\documentclass[aps,preprint,superscriptaddress]{revtex4}
\usepackage{graphics} 
\usepackage{epsfig} 
\usepackage{amsmath}
\usepackage{xcolor}

\usepackage{epstopdf}
\usepackage[caption=false]{subfig}
\usepackage{hyperref}   

\begin{document}
\title{Charged particle dynamics in an elliptically polarized electromagnetic wave and a uniform axial magnetic field}
\author{Shivam Kumar Mishra}
\email{mishrasshivam@gmail.com,shivam.mishra@ipr.res.in}
\affiliation{Institute for Plasma Research, Bhat, Gandhinagar, India, 382428}
\affiliation{Homi Bhabha National Institute, Training School Complex, Anushaktinagar, Mumbai 400094, India}
\author{Sarveshwar Sharma}
\affiliation{Institute for Plasma Research, Bhat, Gandhinagar, India, 382428}
\affiliation{Homi Bhabha National Institute, Training School Complex, Anushaktinagar, Mumbai 400094, India}
\author{Sudip Sengupta}
\affiliation{Institute for Plasma Research, Bhat, Gandhinagar, India, 382428}
\affiliation{Homi Bhabha National Institute, Training School Complex, Anushaktinagar, Mumbai 400094, India}

\date{\today}

\begin{abstract}
An analytical study of the charged particle dynamics in the presence of an elliptically polarized electromagnetic wave and a uniform axial magnetic field, is presented.
 It is found that for $g\omega_{0}/ \omega' = \pm 1$, maximum energy gain occurs respectively for linear and circular polarization; $\omega_{0}$ and $\omega'$ respectively being the cyclotron frequency of the charged particle in the external magnetic field and Doppler-shifted frequency of the wave seen by the particle, and  $g =\pm 1$ respectively correspond to left and right-handedness of the polarization. An explicit solution of the governing equation is presented in terms of particle position or laboratory time, for the specific case of resonant energy gain in a circularly polarized electromagnetic wave.    
These explicit position- or time-dependent expressions are useful for better insight into various phenomena, viz., cosmic ray generation, microwave generation, plasma heating, and particle acceleration, etc.
\end{abstract}
\maketitle
\section{Introduction}
In cyclotron auto resonance particle acceleration (CARPA), the charged particle is subjected to the combined field of transverse electromagnetic (EM) wave along with the homogeneous static axial magnetic field (magnetic field along the wave propagation direction) and it accelerates as a result of self-sustained resonance between the cyclotron motion of the particle (due to external magnetic field) and the oscillation of wave's electric field. The condition for resonance can be achieved by choosing the axial magnetic field and initial momentum of the charged particle so that, initially, the cyclotron frequency of the particle is equal to the Doppler-shifted frequency of the wave seen by the particle. Under this condition, due to the cyclotron motion, the rotation of the charged particle around the wave propagation axis becomes phase-locked with the oscillations of wave's electric field, which results in the charged particle experiencing a DC-component of the electric field (transverse to the wave's propagation direction) and thus gains a net transverse momentum. Simultaneously, the longitudinal force (force along wave propagation direction) due to the wave's magnetic field leads to the transfer of transverse kinetic energy of the particle into it's longitudinal motion such that the Doppler shifted frequency  remains unchanged. This conservation in the Doppler shifted frequency leads to the resonance condition  remaining sustained forever and consequently, the particle gains unbounded energy from the wave. 

CARPA is commonly observed in laboratories as well as naturally occurring phenomena such as plasma heating \cite{Chu_RMP_2004, Hosea_prl_1979,A_Compant_La_Fontaine_1998,ecrh_Bagryansky_2015,PhysRevLett.109.195003,PhysRevE.84.016403,PhysRevLett.80.4681,heating_in_tokamak1,heating_in_fusion_grade_reactor,PhysRevLett.105.165001}, laser-driven particle acceleration \cite{Loeb_prl_1986, Loeb_pra_1986, Loeb_pra_1987, Ride_pre_1995, van_prl_1996, Panofsky_rmp_1999, Vikram_lbap_2013,Vikram_pop_2014,swain2022laser,mishra2022effect,mishra2021exact,Shaar_2023, malik2021analytical, kumar2022diagnostic, kumar2022comparative, kumar12023numerical}, motion of charged particles in the Van-Allen belt \cite{Parker_gp_1961,PARKER19659} and generation of cosmic rays from astrophysical events (like solar flares, the merger of a binary system, a supernova explosion and particle acceleration from the pulsar wind etc.) \cite{hess1963victor,schwarm_2010,salamin2021cosmic,salamin2021cyclotron,salamin2021polarization,salamin2022investigation,strumia2023triggering,belotsky2022mechanisms,salamin2020cosmic,COLEMAN2023102819,MERTEN2021102564,ANCHORDOQUI20191,prajapati2022gravitational,prajapati2022effects}. Particularly, ultra-high energy cosmic rays (UHECRs) with energy greater than $10^{17}-10^{18} eV$ have been detected reaching Earth since 1962 \cite{first_cosmic_ray} and the most energetic particle observed in nature, with energies $(3.2 \pm 0.9) \times 10^{20} eV$ (or $51\pm14 J$) \cite{bird1995detection,larsendetection}.
 The mechanism behind generation of UHECRs  has mainly been referred to acceleration of charged particle by shock waves \cite{Fermi_1979_671_676}. Very recently, the CARPA mechanism (in the simultaneous presence of a super-strong magnetic field and ultra-high intense EM radiation in astrophysical events like merger of a binary system, supernova explosion and pulsar wind) has been proposed as an energy booster for pre-accelerated particles (pre-accelerated via all of the well investigated mechanisms like Fermi-acceleration at shock waves \cite{Fermi_1979_671_676} and magnetic reconnection \cite{sironi2009particle} etc.) \cite{salamin2021cosmic}. Furthermore, it has also be shown that multi-particle simulations (like Particle-in-cell simulation etc) and single particle calculations agree well for energy estimation of cosmic rays in such scenarios \cite{salamin2022investigation,particle_in_cell_simulation,salamin2021cyclotron}. This signifies a fundamental importance for studying a charged particle dynamics in an EM wave and a uniform axial magnetic field. Therefore, in this article, we aim to find the energy, momentum, and position of the particle for above problem as a function of phase of the EM wave (in parametric space), in terms of particle's position coordinates and lab time. In addition to cosmic ray generation, these solutions may also be applied to laboratory experiments, such as plasma heating\cite{Chu_RMP_2004,dhawan2020behaviour} and charged particle acceleration \cite{Milant_ev_2013, DATTOLI2020163310,haraapplication,sabchevski2005cyclotron,
 	jiang2017cyclotron,milant1997cyclotron}.
   
The term auto resonance was first introduced by Andronov \text{et. al.} \cite{Andronov} for the process in which the resonance happens under the action of force induced by the motion of the system itself. The mechanism termed cyclotron auto resonance was discovered by Kolomenski and Lebedev \cite{kolomenskii1962autoresonance,kolomenskii1963self}
and, independently, by Davydovski \cite{DavydovskyVa}.
Roberts and Buchsbaum \cite{Roberts} have studied  particle dynamics pertaining to auto resonance particle acceleration in the presence of circularly polarized EM wave in arbitrary medium. The obtained analysis showed that the resonance condition only sustained for the case of  vacuum ($\mu=1$, here $\mu$ is the refractive index). In this context, an analytical expression which represents asymptotic nature of energy as a function of lab time was derived. The obtained expression conclusively demonstrates that particle acquire an indefinite amount of energy from the wave. On the other hand for $\mu \ne 1$ it was shown that resonance is not sustained and the energy of the charged particle becomes oscillatory in time.
Bourdier and Gond \cite{Bourdier_2,Bourdier_3} studied cyclotron auto resonance in both linearly and circularly polarized EM plane wave using Hamiltonian formalism. Their analysis showed that the motion of charged particles can be accurately described in terms of canonical action-angle variables, indicating a complete integrability of the system. Consequently, an equation describes the evaluation of energy of the particle in resonant case was obtained, which in turn allowed them to rederive the asymptotic solution originally proposed by Roberts and Buchsbaum \cite{Roberts}. 
 This problem for a charged particle in an elliptically polarized wave along a constant axial magnetic field using a potential representing a traveling wave with constant amplitude was solved by R. Ondarza-Rovira \cite{Ondarza-Rovira}. The method of solution allows us to integrate in exact form of relativistic Lorentz force equation which yields analytical expressions for trajectories and velocities of the particle as a function of phase of the EM wave.
 Kong and Liu \cite{Kong}  has derived the position, momentum and energy of the particle as an explicit function of three free parameters namely phase of the EM wave, lab time and longitudinal position of the charged particle. Here derivation of solution represented in terms of phase of the EM wave  follows the similar method as earlier discussed by R. Ondarza-Rovira  \cite{Ondarza-Rovira}. These explicit solutions in terms of longitudinal position and lab time  are not valid for the auto-resonant case. They only apply to a specific set of initial conditions in the non-resonant case (see Appendix \ref{aap2}). 
Many other studies in literature, which have explored the various aspects of cyclotron auto-resonance and discussed their applications can be found in the references \cite{ Chu_RMP_2004,Loeb_pra_1986,Loeb_pra_1987,Loeb_prl_1986,Bao_LiangIEEE1999,Gregory_pf_1990,Egil_Lee_pf_1996,Bourdier_pre_2001,Dattoli,Milant_ev_2013,vikram_thesis,Qian_Bao_Liang1,Ondarza1,Vikram_sagar_1,Ondarza1,M_L_Woolley_1971,jory1968charged}.
In the present study, we initially derived the general solutions for the position, momentum, and energy of the charged particle as a function of phase of the EM wave using the R. Ondarza-Rovira method \cite{Ondarza-Rovira}. Subsequently, we obtained the solution for the resonant case by applying the condition  for cyclotron auto-resonance to the general solutions. Furthermore, we demonstrated that the energy gain by the charged particle depends upon the handedness of the polarization of the EM wave. Next, we examine the limitations associated with exact solutions concerning laboratory time and the longitudinal position of the charged particle, as established by Kong and Liu \cite{Kong}. It is demonstrated that these solutions do not hold true for the resonant case. However, in addressing this specific concern, we have derived explicit solutions for the dynamics of the charged particle (including energy, momentum, and position) as explicit functions of both laboratory time and longitudinal coordinate in the resonant case by using a different approach.   
The flow of the paper is as follows. Section (\ref{sec_2}) contains the analytical solution of the Lorentz force equation for the position, momentum and energy of the charged particle moving in an EM wave and a uniform axial external magnetic field. These obtained solutions are expressed as a explicit function of phase ($\phi$) of the EM wave. In the section (\ref{results}), the results for the energy gain with different polarization of the EM wave has been presented. The explicit position and lab-time dependent solution for circularly polarized EM wave is presented in the Section (\ref{explicite_sol}). Finally, the work is summarized in Section (\ref{Summary}).
\vspace{-1 cc}
\section{Dynamics of the charged particle moving in a transverse EM wave and a uniform axial magnetic field}\label{sec_2}
\subsection{Theory}\label{theory}
Let us consider a charged particle of arbitrary charge and mass moving in a EM wave and a uniform axial magnetic field $B_0$. The governing equation of motion for the charged particle dynamics is given by 
\begin{equation}\label{chap5_eqn1}
	\frac{d\vec{p}}{d t}=\left(\vec{E}+\vec{\beta}\times \vec{B}\right)
\end{equation}
Here the symbols have their usual meanings and the normalizations used are
$t \rightarrow \omega t$, $r \rightarrow k \vec{r}$, $\vec{\beta} = \vec{v}/c$, $\vec{p} \rightarrow {\vec{p}}/{m c}$, $ \vec{E} \rightarrow {q\vec{E}}/{m \omega c }$, $ \vec{B} \rightarrow {q\vec{B}}/{m \omega c }$ ($ {B}_{0} \rightarrow {q{B}_{0}}/{m\omega c }$). Symbols $\omega$ and $k$ respectively representing the frequency and wave vector associated with the EM wave.

The vector potential corresponding to a transverse EM wave can be written as 
\begin{equation}\label{vec_pot_wave}
	\vec{A}= A_x \hat{x} + A_y \hat{y} =  a_0 \delta \cos (\phi)   \hat{x}+ a_0 \sqrt{(1-\delta^2)} g \sin (\phi) \hat{y}
\end{equation}  
Now, vector potential that represents the combined EM field, consisting of the EM wave and a uniform axial magnetic field, can be expressed as follows:
\begin{equation}\label{vec_pot_net}
	\vec{A}_{net}= \left\{ a_0 \delta \cos (\phi) -\frac{1}{2} \left(y \omega_{0}\right)\right\}  \hat{x}+ \left\{a_0 \sqrt{(1-\delta^2)} g \sin (\phi) +\frac{1}{2} \left( x \omega_{0}\right)\right\}\hat{y}
\end{equation}       
Here, $\phi= t-z + \phi_0$ ($\phi_0$ is the phase of the EM wave at $t=0$); and $x$, $y$ and $z$ are the components of the position vector $``r"$ along the unit vectors $\hat{x}$, $\hat{y}$ and $\hat{z}$ (right-handed Cartesian coordinates system) respectively. In addition,  $\delta  \in  [0, 1]$, where $\delta = 0, 1$ and $\delta = 1/\sqrt{2} $ correspond to linear and circular polarization respectively; and $g = \pm 1$ respectively correspond to  left and right-handedness of polarization.

%
%

Using $\vec{E} = - \partial \vec{A}_{net} / \partial t$ and $\vec{B}= \vec{\nabla} \times \vec{A}_{net}$, the transverse and longitudinal components of equation (\ref{chap5_eqn1}) may respectively be written as
\begin{equation}\label{chap5_eqn2}
\hspace{0.6cm}	\frac{d\vec{p}_{\perp}}{d t}= -\left\{\frac{d{\vec{A}}}{d{t}}+\left(\hat{z} \times \vec{\beta} \right) \omega_0 \right\}
\end{equation} 
\begin{equation}\label{chap5_eqn3}
	\frac{d{p_z}}{d t}= \left\{\beta_x \frac{\partial{A_x}}{\partial{z}}+\beta_y\frac{\partial{A_y}}{\partial{z}} \right\}
\end{equation}
Here, $\omega_{0}$ is the cyclotron frequency corresponding to axial uniform magnetic field $B_{0}$.
The rate of change of energy of the charged particle (by taking dot product of equation (\ref{chap5_eqn1}) with $\vec{\beta}$) is given by

\begin{equation}\label{chap5_eqn4}
	\frac{d{\gamma}}{d t}=- \left\{\beta_x \frac{\partial{A_x}}{\partial{t}}+\beta_y\frac{\partial{A_y}}{\partial{t}} \right\}
\end{equation}

Where $\vec{p}_{\perp} =p_{x} \hat{x} + p_{y} \hat{y}$ and $p_z$ is the momentum of the charged particle in the direction of propagation of the wave. 
It can be seen from equation (\ref{chap5_eqn2}) that the transverse momentum of the particle has an exact solution in terms of the phase of the EM wave, which is given by
\begin{equation}\label{chap5_eqn5}
	\vec{P}_{\perp0}=\vec{p}_{\perp}+ \vec{A}(\phi)- \left(\vec{r} \times \hat{z}   \right) \omega_0
\end{equation}
Where $\vec{P}_{\perp0}$  ($=\vec{p}_{\perp 0}+ \vec{A}(\phi_0)$)  is a constant of motion and $\vec{p}_{\perp 0}$ is the value of $\vec{p}_{\perp}$ at $t=0$.
Here, the initial condition are chosen such that the charged particle with arbitrary initial momentum is placed at the origin ($\vec{r}_{0} = 0$), observing an arbitrary phase of a polarized EM wave. 
Now, by subtracting the equation (\ref{chap5_eqn3}) from the equation (\ref{chap5_eqn4}) and integrating, one can write:
\begin{equation}\label{delta}
 (\gamma-{p_z})= \Delta(=\gamma_{0}-p_{z0}) 
\end{equation}
 Here $\Delta$ turns out to be a constant of motion (where $\vec{r}_{0}$, $\vec{p}_{\perp0}$, $p_{z0}$ and $\gamma_{0}$ are the initial position, initial transverse momentum, initial longitudinal momentum and initial energy of the charged particle, respectively).
 
Now, employing the expression $ \vec{p}= \Delta ({d \vec {r}}/{d \phi})$, equation (\ref{chap5_eqn5}) can be expressed in the following manner:
\begin{equation}\label{chap5_eqn6}
	\frac{d^2 x}{d \phi^2}+{\Omega_0}^2 x=\left(w_1-\Omega_0 g w_2\right) \sin \phi + \frac{\Omega_{0} P_{y0}}{ \Delta}
\end{equation}
\begin{equation}\label{chap5_eqn7}
	\frac{d^2 y}{d \phi^2}+{\Omega_0}^2 y=\left(w_1 \Omega_0- g w_2\right) \cos \phi - \frac{\Omega_{0} P_{x0}}{\Delta}
\end{equation}
Here $\Omega_0={\omega_0}/{\Delta}$, $w_1={a_0 \delta}/{\Delta}$, $w_2= {a_0  \sqrt{1-\delta^2}}/{\Delta}$; and $P_{x0}$ and $P_{y0}$ are the components of the $\vec{P_{0}}$ in the $x$ and $y$ direction respectively. 
The equations (\ref{chap5_eqn6}) and (\ref{chap5_eqn7}) resemble forced oscillators and the condition ${\Omega_0}^2 =1$ 
gives the resonance condition.
Since $\Delta$ is a constant of motion, hence resonance condition is sustained throughout the motion. In the resonant condition particle gets phase locked with the wave and gains unbounded energy. On the other hand for the non-resonant case (${\Omega_0}^2 \neq 1$) energy becomes oscillatory. 
It is important to note that ${\Omega_0} = 1$ leads to the condition $\omega_{0} = \Delta \omega (\equiv \omega')$, and hence, it results that in the resonant case, the cyclotron frequency of the particle equals the Doppler-shifted frequency of the wave as seen by the particle.
%
%
\subsection{Explicit solution of the governing equation in terms of phase of EM wave}\label{non_reso_particle}
The exact solution of the equations (\ref{chap5_eqn6}) and (\ref{chap5_eqn7}) for the case of $\phi_{0}=0$ is given by 
\begin{equation}\label{chap5_eqn8}
\hspace{-0.5 cc}	x = \frac{P_{y0}}{\Delta \Omega_{0}}\left(1 -  cos(\Omega_{0} \phi)\right) + \left\{\frac{p_{x0}}{\Delta \Omega_{0} } - \frac{W_{1}}{\Omega_{0}(\Omega_{0}^{2}-1)}\right\} \sin(\Omega_{0} \phi) + \frac{W_{1}}{(\Omega_{0}^{2}-1)} sin(\phi)
\end{equation}\\
\begin{equation}\label{chap5_eqn9}
	\begin{split}
	y=-\frac{P_{x0}}{\Delta \Omega_{0}}
		+ \frac{p_{y0}}{\Delta \Omega_{0}} sin(\Omega_{0} \phi)  + \left\{\frac{P_{x0}}{\Omega_{0} \Delta} - \frac{W_{2}}{(\Omega_{0}^{2}-1)}\right\} \cos(\Omega_{0} \phi) + \frac{W_{2}}{(\Omega_{0}^{2}-1)} 
		cos(\phi)
	\end{split}
\end{equation}
Here, $W_{1} = (w_{1}-  g \Omega_{0} w_{2})$ and $W_{2} = (\Omega_{0} w_{1} -g  w_{2})$. The governing equation for the motion of the charged particle along the wave propagation direction is given by
\begin{equation}\label{chap5_eqn10}
	\frac{d^2 z}{d \phi^2}= w_1  \sin{\phi} \frac{d x}{d \phi} -w_2 g \cos{\phi} \frac{d y}{d \phi} 
\end{equation}
The exact solution of equation (\ref{chap5_eqn10}) for longitudinal momentum and position is given by  
\begin{equation}\label{chap5_eqn11}
	\begin{split}
		\hspace {-0.5 cc } p_{z}=p_{z0} -\frac{\alpha_{0} sin(\Omega_{0}+1)\phi}{2 (\Omega_{0}+1)}  + \frac{\beta_{0} sin(\Omega_{0}-1)\phi}{2 (\Omega_{0}-1)}  -\Delta\Bigg\{ 
		\alpha_{1} \frac{cos(\Omega_{0}+1)\phi}{(\Omega_{0}+1)}+ \alpha_{2}  \frac{cos(\Omega_{0}-1)\phi}{(\Omega_{0}-1)}\\+\frac{(w_{1}^{2}-w_{2}^{2})}{4(\Omega_{0}^{2}-1)}cos2\phi+ \alpha_{3} \Bigg\}
	\end{split}
\end{equation}
\begin{equation}\label{chap5_eqn12}
	\begin{split}
	\hspace{-1.4 cc}	z= \alpha_{4} +\frac{\alpha_{0} cos(\Omega_{0}+1)\phi}{2 \Delta (\Omega_{0}+1)^{2}} - \frac{\beta_{0}  cos(\Omega_{0}-1)\phi}{2 \Delta (\Omega_{0}-1)^{2}}   
		- \alpha_{1}  \frac{sin(\Omega_{0}+1)\phi}{(\Omega_{0}+1)^{2}} - \alpha_{2} \frac{sin(\Omega_{0}-1)\phi}{(\Omega_{0}-1)^{2}}-\\ \frac{(w_{1}^{2}-w_{2}^{2})}{8(\Omega_{0}^{2}-1)}sin2\phi  + \alpha_{5}{\phi}
	\end{split}
\end{equation}
Here
$$ \hspace{-15 cc} \alpha_{0}=\big(w_{1} P_{y0}+w_{2} g p_{y0}\big), \beta_{0}=\big(w_{1} P_{y0}-w_{2} g p_{y0}\big)$$
$$\hspace{-3.7 cc} \alpha_{1} = \frac{\big(w_{1} p_{x0}+w_{2} g P_{x0}\big)}{2 \Delta}+\frac{1}{2 (\Omega_{0}^{2}-1)}\Big\{\big(-w_{1}^{2}+w_{2}^{2} \Omega_{0}\big) + w_{1} w_{2} g\Omega_{0} (1-\Omega_{0})\Big\}$$
$$\hspace{-3.5 cc} \alpha_{2} = \frac{\big(-w_{1} p_{x0}+w_{2} g P_{x0}\big)}{2 \Delta}+\frac{1}{2 (\Omega_{0}^{2}-1)}\Big\{\big(w_{1}^{2}+w_{2}^{2} \Omega_{0}\big) - w_{1} w_{2} g\Omega_{0} (1+\Omega_{0})\Big\}$$
$$\hspace{-2.5 cc}\alpha_{3}=-\Bigg\{\frac{ \alpha_{1} }{{\Omega_{0} +1}} +  \frac{\alpha_{2}} {\Omega_{0} -1} +\frac{(w_{1}^{2}-w_{2}^{2})}{4(\Omega_{0}^{2}-1)} \Bigg\}, \alpha_{4} = \frac{ -\alpha_{0}}{2 \Delta \left(\Omega_{0}+1\right)^{2}} +  \frac{ \beta_{0}}{2 \Delta \left(\Omega_{0}-1\right)^{2}}$$
$$\hspace{-16 cc}\alpha_{5}=\frac{p_{z0}}{\Delta} + \frac{ \alpha_{1} }{{\Omega_{0} +1}} +  \frac{\alpha_{2}} {\Omega_{0} -1}  + \frac{\big(w_{1}^{2} - w_{2}^{2}\big)}{4({\Omega_{0}^{2} -1})}$$
The expression (\ref{chap5_eqn11}) represents the longitudinal momentum of the particle. Whereas expressions (\ref{chap5_eqn8}), (\ref{chap5_eqn9}) and (\ref{chap5_eqn12})  represent the $x$, $y$ and $z$ components of the trajectory of the particle respectively. 
\subsection{ Explicit solution of the governing equation in terms of phase of EM wave for resonant case}\label{reso_particle}
For resonant phase locking in the case of circularly polarized EM wave, the  rotation of electric field vector as seen by charged particle, must occur at the same speed and in the same direction as the particle's cyclotron rotation caused by the external magnetic field. For this, it is necessary for $g$ and $\Omega_{0}$ to have opposite signs.  
This observation can be validated by examining the expressions of $(w_1-\Omega_0 g w_2)$ and $(w_1 \Omega_0- g w_2)$ respectively in equations (\ref{chap5_eqn6}) and (\ref{chap5_eqn7}) in relation to $w_{1}$ and $w_{2}$. By doing so, it becomes apparent that the right-hand side of the force harmonic equations (equations (\ref{chap5_eqn6}) and (\ref{chap5_eqn7})) can only persist when the condition $\Omega_{0} = -g$ is satisfied.

Thus in the resonant case, for a circularly polarized EM wave, negative charge gains energy only from the left handed polarization case ($g=+1$, wave with negative helicity) and positive charge gains energy only from the right handed polarization  case ($g=-1$, wave with positive helicity). This same idea for energy gain can further be used for the case of elliptically polarized EM wave in the following manner. 

The vector potential defined in equation (\ref{vec_pot_wave}) may be expressed as

\begin{equation}\label{vec_pot_el}
	\vec{A}=\frac{\delta+\sqrt{(1-\delta^2)}}{2} a_0
	\Big\{\cos (\phi)   \hat{x}+  g \sin (\phi) \hat{y}\Big\} 
	+  \frac{\delta-\sqrt{(1-\delta^2)}}{2} a_0 \Big\{\cos (\phi)   \hat{x}-  g \sin (\phi) \hat{y}\Big\}
\end{equation}
Equation (\ref{vec_pot_el}) demonstrated that the vector potential corresponding to the elliptically polarized EM wave can be expressed as the vector sum of two circularly polarized waves having different amplitude and opposite helicity (if the magnitude of amplitudes of these two circularly polarized waves are same then vector potential represents linearly polarized EM wave  and for the case of circularly polarized EM wave ($\delta=1/ \sqrt{2}$) only first term on right hand side of equation (\ref{vec_pot_el}) survive). Now, if we define 
\begin{equation}\label{vec_pot_el_1}
	\vec{A}_{1}=\frac{\delta+\sqrt{(1-\delta^2)}}{2} a_0
	\Big\{\cos (\phi)   \hat{x}+  g \sin (\phi) \hat{y}\Big\} 	
\end{equation}
and
\begin{equation}\label{vec_pot_el_2}
	\vec{A}_{2}=  \frac{\delta-\sqrt{(1-\delta^2)}}{2} a_0 \Big\{\cos (\phi)   \hat{x}-  g \sin (\phi) \hat{y}\Big\}
\end{equation}
\
Then, for the cyclotron auto resonance in elliptically polarized EM wave, the choices $\Omega_{0}=-g$  and $\Omega_{0}=g$  respectively leads to phase-locking with the wave characterized by vector potentials $\vec{A}_{1}$ and $\vec{A}_{2}$; and result in unbounded energy gain by the particle (in either of the cases, the energy gain associated with the EM wave which does not get phase locked exhibits an oscillatory nature). In contrast to a circularly polarized EM wave, where resonance only occurs for $\Omega_{0}=-g$, in the case of an elliptically polarized EM wave ($\delta \ne 1/\sqrt{2}$), resonance occurs for both choices of $\Omega_{0}=-g$ and $\Omega_{0}=g$. 

The expressions for the momentum, energy, and position of a positive charged particle in the resonance case can be determined by taking the limit $\Omega_{0} \rightarrow 1$ in the expressions given in the sub-section (\ref{non_reso_particle}). Consequently, we obtain the following results:
\begin{equation}\label{px_reso}
	\hspace{-2.5 cm}{p}_{x}	={P}_{x0}-w_{1}\Delta cos(\phi)+  {p_{x0}} \left(-1+cos(\phi)\right) + {p_{y0}} sin(\phi) + \frac{(w_{1}-g w_{2} )\Delta}{2} \phi sin(\phi) 
\end{equation}
\begin{equation}\label{py_reso}
	\hspace{-3.1 cm} {p}_{y}	={P_{y0}}  cos(\phi) - {p_{x0}} sin(\phi) - \frac{\left(w_{1}+g w_{2} \right)\Delta}{2}sin(\phi) + \frac{\left(w_{1}-g w_{2} \right)\Delta}{2} \phi cos(\phi)   
\end{equation}
\begin{equation}\label{pz_reso}
	\begin{split}
\hspace{-2.0 cm}	p_{z}= p_{z0}- \frac{\alpha_{0}}{2} sin(2\phi)  + \frac{(w_{1}p_{x0} +  w_{2}g P_{x0})}{4} (1-cos(2\phi))  - \Delta \Bigg\{ \frac{(w_{1}-gw_{2})^{2}}{16} cos(2\phi) \\+\frac{\big(w_{1}^{2} - w_{2}^{2} \big)}{8} \phi sin(2\phi)  - \frac{(w_{1}-gw_{2})^{2}}{16} \Big( 1+ 2 \phi^{2} \Big) \Bigg\} + \frac{\beta_{0}}{2} \phi
	\end{split}
\end{equation}
\begin{equation}\label{gamma_reso}
	\begin{split}
\hspace{-2.1 cm}	\gamma= \gamma_{0}- \frac{\alpha_{0}}{2} sin(2\phi)  + \frac{(w_{1}p_{x0} +  w_{2}g P_{x0})}{4} (1-cos(2\phi))  - \Delta \Bigg\{ \frac{(w_{1}-gw_{2})^{2}}{16} cos(2\phi) \\+\frac{\big(w_{1}^{2} - w_{2}^{2} \big)}{8} \phi sin(2\phi)  - \frac{(w_{1}-gw_{2})^{2}}{16} \Big( 1+ 2 \phi^{2} \Big) \Bigg\} + \frac{\beta_{0}}{2} \phi
	\end{split}
\end{equation}
\begin{equation}\label{x_reso}
\hspace{-4.3 cm} x= \frac{P_{y0}}{\Delta}\left(1 -  cos(\phi)\right) + \frac{p_{x0}}{\Delta} sin(\phi) + \frac{\left(w_{1}-g w_{2} \right)}{2} \Big(-\phi cos(\phi) + sin(\phi)\Big)
\end{equation}
\begin{equation}\label{y_reso}
\hspace{-6.5 cm}	y= \frac{p_{x0}}{\Delta} \left(-1+cos(\phi)\right) + \frac{p_{y0}}{\Delta} sin(\phi) + \frac{(w_{1}-g w_{2} )}{2} \phi sin(\phi)
\end{equation}
\begin{equation}\label{z_reso}
	\begin{split}
\hspace{-0.6 cm}	z= \frac{p_{z0}}{ \Delta} \phi+ \frac{\alpha_{0}}{8 \Delta} \left( 1- cos(\phi) \right)  + \frac{(w_{1}p_{x0} +  w_{2}g P_{x0})}{8\Delta} (2\phi-sin(2\phi)) - \frac{(w_{1}-gw_{2})^{2}}{32} {sin(2\phi)}\\ - \frac{\big(w_{1}^{2} - w_{2}^{2} \big)}{32 } \Big(-2\phi cos(2\phi) + sin(2\phi)\Big)  + \frac{(w_{1}-gw_{2})^{2}}{16} \Big( \phi+ \frac{2}{3} \phi^{3} \Big)  + \frac{\beta_{0}}{4 \Delta} \phi^{2}
	\end{split}
\end{equation}
The equations (\ref{px_reso}), (\ref{py_reso}) and (\ref{pz_reso}) respectively represent the x, y and z components of the momentum of the particle for the resonant case; and expression (\ref{gamma_reso}) represents energy of the charged particle. It shows that the energy of the charged particle increases monotonically with time. The equations (\ref{x_reso}), (\ref{y_reso}) and (\ref{z_reso}) respectively represent the x, y and z components of the trajectory of the particle. 

The above expressions (equation (\ref{chap5_eqn8})-(\ref{chap5_eqn12}) and equation (\ref{px_reso})-(\ref{z_reso})) are derived for $\phi_{0}=0$ case. The expressions for arbitrary choice of $\phi_{0}$ for both non-resonant and resonant cases are given in the Appendix-(\ref{app1}). 
%
\section{Results}\label{results} 
\begin{figure} 
	\includegraphics[width= 16 cm, height=10cm]{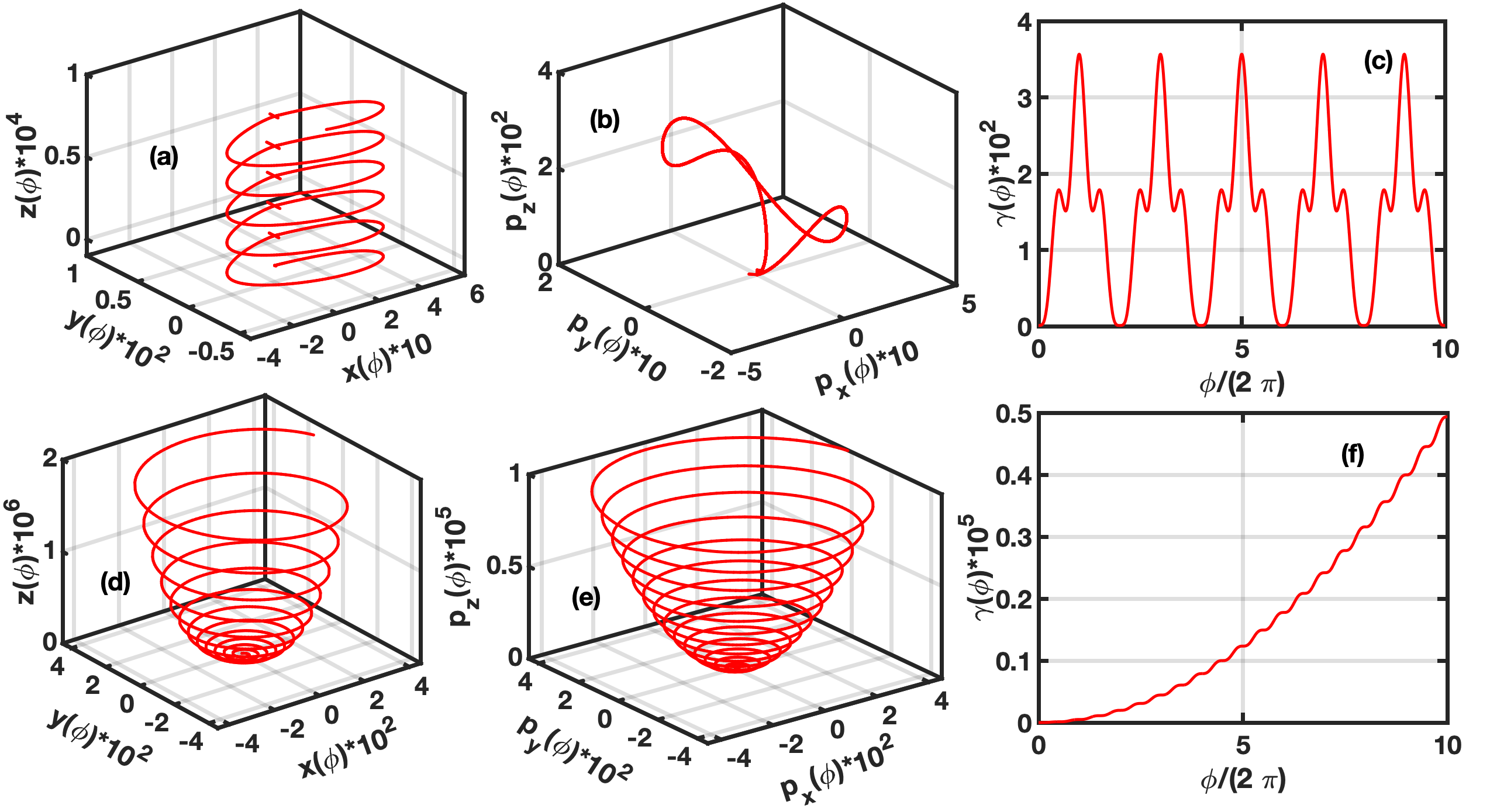}
	\caption{Figure represents the dynamics of the charged particle in linearly polarized EM wave ($\delta=1$) for both non-resonant and resonant cases. Sub-plots a-c respectively representing the evolution of trajectory, momentum and energy of the particle with time ($\phi$) for non-resonance case ($\Omega_{0}=0.5$). Whereas, sub-plots (d)-(f) respectively representing the evolution of trajectory, momentum and energy of the particle with time ($\phi$) for the resonance case. These graphs are plotted for positive charged particle initially at rest interacting with the EM wave with $a_{0}=10$ and $g = -1$.}
	\centering
	\label{chap_5_fig_1}
\end{figure}

\begin{figure}
	\includegraphics[width= 16 cm, height=10cm]{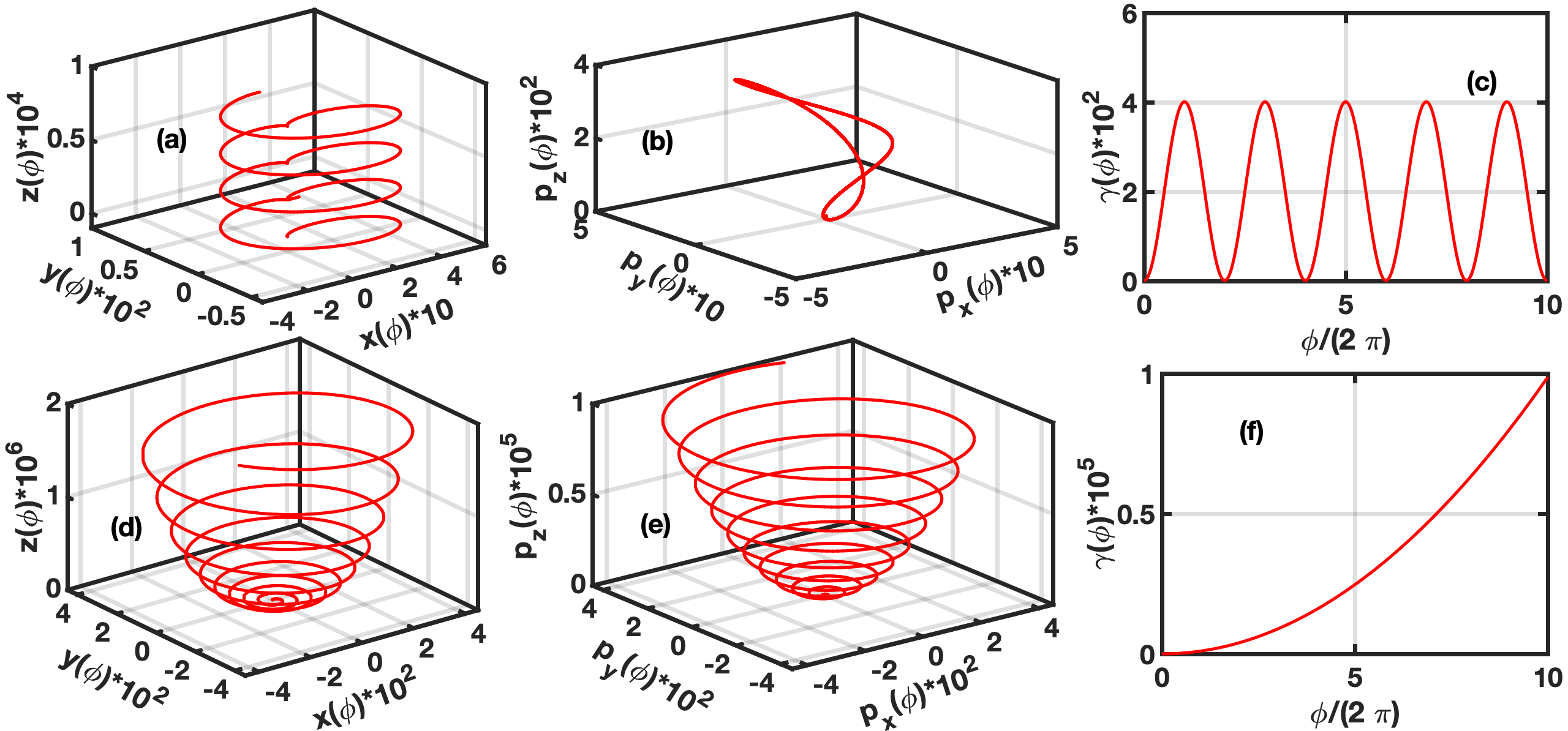}
	\caption{Figure represents the dynamics of the charged particle in circularly polarized EM wave $\left(\delta=1/\sqrt{2}\right)$ for both non-resonant and resonant cases. These  sub-plots (a)-(c) respectively representing the evolution of trajectory, momentum and energy of the particle with time ($\phi$) for non-resonance case ($\Omega_{0}=0.5$). Whereas, sub-plots (d)-(f) respectively representing the evolution of trajectory, momentum and energy of the particle with time ($\phi$) for the resonance case. These graphs are plotted for positive charged particle initially at rest interacting with the EM wave with $a_{0}=10$ and $g = -1$.}
	\centering
	\label{chap_5_fig_2}
\end{figure}

\begin{figure}
	\includegraphics[width= 16 cm]{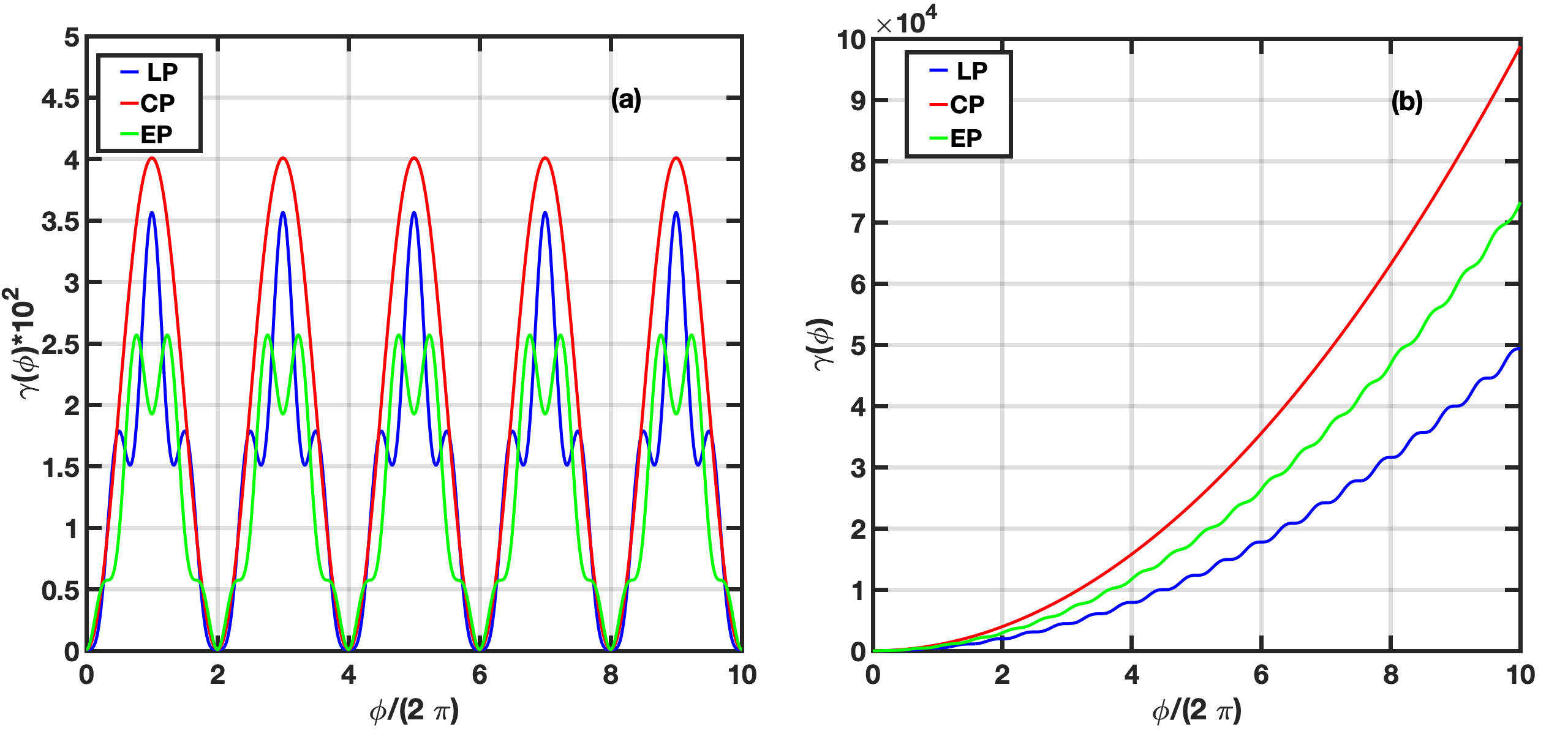}
	\caption{Figure represents the comparison of energy gain by the particle in different polarization of the EM wave. The sub-plots (a) and (b) respectively representing the energy gain for non-resonance case ($\Omega_{0}=0.5$) and resonant case. Here blue, red and green curves respectively representing the  linear ($\delta=1$), circular $\left(\delta=1/\sqrt{2}\right)$  and elliptically ($\delta=1/4$) polarized EM wave.   These graphs are plotted for positive charged particle initially at rest interacting with the EM wave with $a_{0}=10$ and $g = -1$.}
	\centering
	\label{chap_5_fig_3}
\end{figure}

\begin{figure}
	\centering
	\includegraphics[width=16.5 cm, height=6cm]{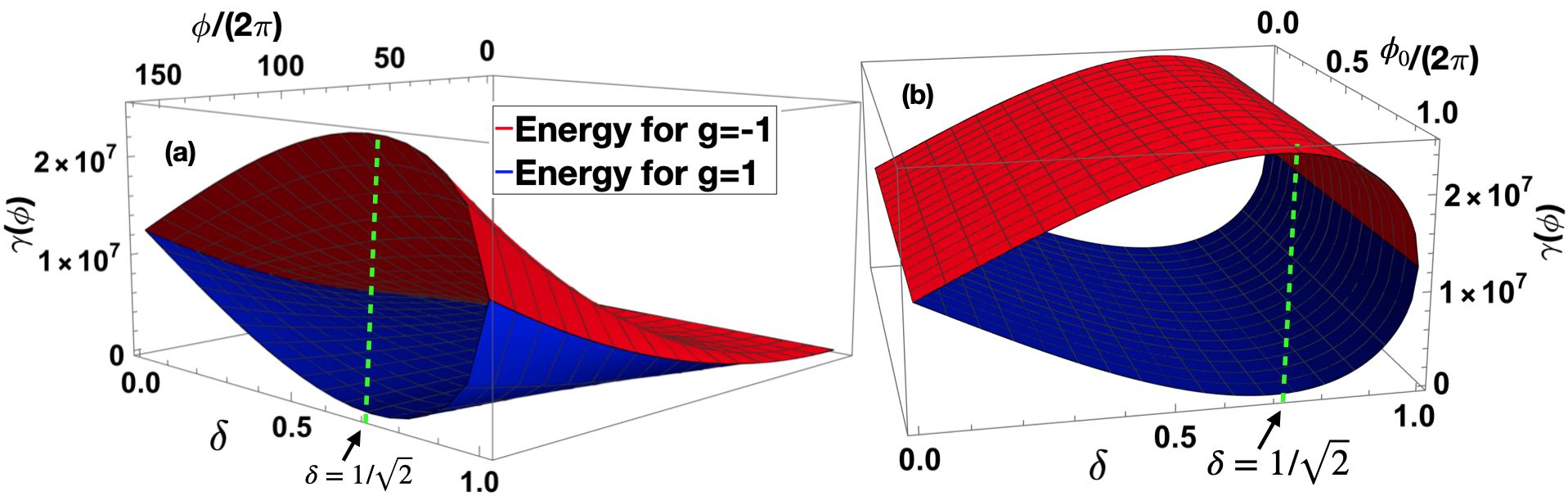}
	\caption{Figure represents the energy of the charged particle as a function of $\phi$, $\delta$, $g$ and $\phi_{0}$  for resonant case ($\Omega_{0} = 1$). Sub-plots (a) and (b) respectively representing the evolution energy with $\phi$ ( for $\phi_{0}=0$) and $\phi_{0}$ (for $\phi=150$)  for different values of $\delta$ and $g$. Colors blue and red respectively correspond to $g=1$ and $g=-1$. These graphs are plotted for positive charged particle initially at rest interacting with the EM wave with $a_{0}=10$.}
	\label{energy_with_delta_g}
\end{figure}

In this section, the dynamics of the charged particle has been analyzed by plotting our analytical results for non-resonant and resonant cases respectively.
Figures \ref{chap_5_fig_1}, \ref{chap_5_fig_2} and \ref{chap_5_fig_3} respectively represent the dynamics of the charged particle for linearly polarized EM wave ($\delta=1$), circularly polarized EM wave $\left(\delta=1/\sqrt{2}\right)$ and comparison of the energy gain with different polarization of the EM wave. For these plots, the set of initial condition chosen to be  $\vec{p}_{0}=0$ (particle starts at rest), $a_{0}=10$ and $g=-1$. 
In the figures \ref{chap_5_fig_1} and \ref{chap_5_fig_2},  (a)-(c) respectively represent the evolution of trajectory, momentum and energy of the particle with time ($\phi$) for non-resonant case. Whereas, (d)-(f) respectively represent the evolution of trajectory, momentum and energy of the particle with time ($\phi$) for the resonant case. 
In the figure \ref{chap_5_fig_3}, (a) and (b) represent the comparison of energy gain for  different polarization in non-resonant and resonant cases, respectively. The curves blue, red and green in figure \ref{chap_5_fig_3} corresponds to the linearly ($\delta=1$), circularly $\left(\delta=1/\sqrt{2}\right)$ and elliptically ($\delta=1/4$) polarized EM wave respectively. Next, figure \ref{energy_with_delta_g} represents the energy of the charged particle as a function of $\phi$, $\delta$, $g$ and $\phi_{0}$  for resonant case ($\Omega_{0} = 1$). 
Sub-plots (a) and (b) respectively representing the evolution energy with $\phi$ and $\phi_{0}$  for different values of $\delta$ and $g$. The blue and red colors respectively correspond to $g=1$ and $g=-1$. These graphs are plotted for positive charged particle initially at rest interacting with the EM wave with $a_{0}=10$.    

In the sub-figure \ref{chap_5_fig_1}(a), red curve represents a helical path for the evolution of trajectory for the non-resonant particle ($\Omega_{0}=0.5$).
It clearly indicates that both the radius of gyration and pitch are the constant of motion. This implies that the transverse motion of the particle is oscillatory and it drifts along the longitudinal direction. This is also apparent in the recurring pattern observed in the momentum of the particle, as shown in sub-figure \ref{chap_5_fig_1}(b). The average over the transverse momentum, vanishes, while the non-zero average of the longitudinal motion causes the charged particle to drift in the positive z-direction. The corresponding energy depicted in sub-figure \ref{chap_5_fig_1}(c) exhibits a periodic pattern and average energy correspond to the drift velocity of the particle. 
In contrast to this, for the resonant case ($\Omega_{0}=1$), particle trajectory in configuration and momentum space (shown in sub-figure  \ref{chap_5_fig_1}(d) and \ref{chap_5_fig_1}(e)) shows a helical path with increasing pitch and radius of gyration.
This describes a monotonic increase in both the transverse and longitudinal momentum of the charged particle, resulting in the continuous energy gain by the particle from the wave (shown in sub-figure  \ref{chap_5_fig_1}(f)).
Here, it is essential to note that the increase in longitudinal momentum occurs at a significantly faster rate compared to the gain in transverse momentum for the particle. As a result, the charged particle's motion tends to follow an almost straight line trajectory (see sub-figure \ref{chap_5_fig_1}(d)). The energy gain by the particle is primarily due to the increase in its longitudinal momentum.
This unbounded energy gain in the resonant case happens due to phase locking between the  cyclotron motion of the particle and oscillation in the electric field vector of the EM wave seen by the particle, which is discussed in the section (\ref{theory}). The dynamics of the charged particle in circularly polarized EM wave, as shown in the figures \ref{chap_5_fig_2}, is qualitatively same as that in linearly polarized  EM wave.   

Next, the energy gain with different polarization of the EM wave for non-resonant and resonant cases is shown in the sub-figures \ref{chap_5_fig_3}(a) and \ref{chap_5_fig_3}(b), respectively. The blue, red, and green curves in the sub-figure \ref{chap_5_fig_3}(a) represent the oscillatory behavior of the energy for all the polarizations in the non-resonant case. The average energy turns out to be of the order of $a^{2}_{0}$.
While for the resonant case, (shown in sub-figure \ref{chap_5_fig_3}(b)) energy of the charged particle increases continuously and net instantaneous energy gain is following a descending order from circular to linear polarization. 
This dependency of energy gain on phase and polarization of the EM wave may be understood as follows. The leading order term $\left\{{\Delta (w_{1}-gw_{2})^{2}}/{8} \right\}\phi^{2} \left(=(a_{0}^{2}/8\Delta) (\delta-g\sqrt{1-\delta^{2}})^{2}\phi^{2}\right)$ in equation (\ref{gamma_reso}) represents quadratic dependency of energy on phase of the EM wave and leads to unbounded energy gain. 
On the other hand, the dependency of energy gain on polarization of the EM wave may be understood by calculating the extremum values of the term $\left(\delta-g\sqrt{1-\delta^{2}}\right)^{2}$.
%
This shows that for the choice of $g=-1$, the term $\left(\delta-g\sqrt{1-\delta^{2}}\right)^{2}$ attains a maximum value of $``2"$ when $\delta=1/\sqrt{2}$ and a minimum value of $``1"$ when $\delta=0,1$. The energy gain for arbitrary polarization can be expressed in terms of the energy gain in a linear or circularly polarized EM wave using the relationship $\gamma_{\delta} \approx \left(\delta+\sqrt{1-\delta^{2}}\right)^{2}\gamma_{l} \approx 1/2 \left(\delta+\sqrt{1-\delta^{2}}\right)^{2}\gamma_{c}$ (Here $\gamma_{l}$, $\gamma_{c}$ and $\gamma_{\delta}$ are energy of the particle for linear, circular and  arbitrary polarization of the EM wave respectively). This clearly explains the difference in energy gain between different polarization of the EM wave, as illustrated in sub-figure \ref{chap_5_fig_3}(b). This same energy gain can also be understood by looking at the dynamics of the charged particle in context to the vector potentials $\vec{A}_{1}$ and $\vec{A}_{2}$. 
As discussed in sub-section (\ref{reso_particle}) that for $\Omega_{0} = 1$ and $g=-1$ the cyclotron rotation of the charged particle leads to phase-locking with the wave characterized by vector potential $\vec{A}_{1}$. The energy gain which is approximately proportional to the amplitude square corresponds to the same proportionality constant $\left(\delta+\sqrt{1-\delta^{2}}\right)^{2}$. The difference in energy gain with different polarization depends on division of net intensity of the EM wave into the intensities of waves corresponding to the vector potentials $\vec{A}_{1}$ and $\vec{A}_{2}$. For example, in case of circularly polarized EM wave $\left(\delta=1/\sqrt{2}\right)$ only $\vec{A}_{1}$ exists ($\vec{A}_{2} = 0$) and term $\left(\delta+\sqrt{1-\delta^{2}}\right)^{2}$ attains a maximum value equal to $``2"$. On the other hand, the intensity of linearly polarized EM wave ($\delta=0$ or $1$) is equally divide into the intensities corresponding to $\vec{A}_{1}$ and $\vec{A}_{2}$; and term $\left(\delta+\sqrt{1-\delta^{2}}\right)^{2}$ attains a minimum value equal to $``1"$.  For any other polarization the term $\left(\delta+\sqrt{1-\delta^{2}}\right)^{2}$ achieves value between $1$ and $2$. Furthermore, it has also discussed in sub-section (\ref{reso_particle}) that except for circular polarization for any other polarization the condition $\Omega_{0} = 1$ and $g=1$ also leads to phase locking of charged particle dynamics with wave associated with vector potential $\vec{A}_{2}$. In these scenarios the energy gain is maximum for linear polarization $\left((\delta-\sqrt{1-\delta^{2}})^{2}=1\right)$ and minimum for circular polarization $\left((\delta-\sqrt{1-\delta^{2}})^{2}=0\right)$ (That is also evident from the leading order term $\left((a_{0}^{2}/8\Delta) (\delta-g\sqrt{1-\delta^{2}})^{2}\phi^{2}\right)$ of equation (\ref{gamma_reso})); which is clearly shown in the sub-figure \ref{energy_with_delta_g}(a).  For the case $g \Omega = 1$, the energy gain for arbitrary polarization can also be expressed in terms of the energy gain in a linear or circularly polarized EM wave using the relationship $\gamma_{\delta} \approx \left(\delta - \sqrt{1-\delta^{2}}\right)^{2}\gamma_{l} \approx \frac{1}{2} \left(\delta - \sqrt{1-\delta^{2}}\right)^{2}\gamma_{c}$. Thus, the net resonant energy gain described by the leading order term for arbitrary polarization and  handedness can be expressed as  
\begin{equation}\label{dif_delta} 
	\gamma_{\delta} \approx \frac{1}{2}
	\begin{cases}
		 \left(\delta+\sqrt{1-\delta^{2}}\right)^{2}\gamma_{c}, & \text{for }  g \Omega = -1 \\	
		
		 \left(\delta-\sqrt{1-\delta^{2}}\right)^{2}\gamma_{c},  & \text{for } g \Omega = 1
	\end{cases}
\end{equation}
It can be clearly shown from the equation (\ref{dif_delta}) that the net energy gain described for arbitrary polarization is always larger for $g \Omega = -1$ case compared to $g \Omega = 1$ case.  
Next, the sub-figure \ref{energy_with_delta_g}(b) it is also clear that the qualitative as well as quantitative nature of the energy gain is independent of the choice of initial phase ($\phi_{0}$) of the EM wave.

{ Thus, from the above discussion, it is clearly evident that the dynamics and the net resonant energy gain by the particle for arbitrary polarization of the EM wave can be clearly understood in terms of dynamics of the charged particle in the presence of circularly polarized EM wave. Therefore, in the next section, derivation of explicit solution for energy, momentum and position of the charged particle in terms of particle position and laboratory time has been discussed only for circularly polarized EM wave.}        
\section{EXPLICIT SOLUTION OF the EQUATION OF MOTION IN TERMS OF particle position and LABORATORY TIME}\label{explicite_sol}
In this section a new approach has been used to calculate the particle's trajectory, momentum and energy as explicit functions of laboratory time and particles position co-ordinate for resonant particle in circularly polarized EM wave and an axial magnetic field, which is as follows:

The vector potential for the EM wave in right-handed cylindrical polar coordinates $(\rho,\theta,z)$ can be written as
\begin{equation}\label{chap5_eqn16}
	\vec{A}= a_{0} cos(\phi-g \theta) \hat{\rho} + a_{0} g sin(\phi-g \theta) \hat{\theta}
\end{equation}
The electric and magnetic fields for the EM wave in the cylindrical polar coordinates system are respectively given by
$\vec{E}= a_{0} sin(\phi-g \theta) \hat{\rho} - a_{0} g cos(\phi-g \theta) \hat{\theta}$ \  \text{and}\ $ \vec{B}= a_{0}g cos(\phi-g \theta) \hat{\rho} + a_{0} sin(\phi-g \theta) \hat{\theta}$. Here $\hat{\rho}$, $\hat{\theta}$ and $\hat{z}$ are the unit vectors along ${\rho}$, $\theta$ and $z$ directions, respectively. Using these expressions of the electric and magnetic fields in the equation (\ref{chap5_eqn1}), the transverse and longitudinal components of the force may respectively be written as
\begin{equation}\label{chap5_eqn17}
	\begin{split}
		\frac{d \vec{p}_{\perp}}{d t}=  \left\{a_{0}  sin(\phi-g \theta) - a_{0}\beta_{z}  sin(\phi-g \theta) +  \omega_{0}\beta_{\theta} \right\} \hat{\rho} - \big\{ a_{0} g  cos(\phi-g \theta) -  a_{0} g \beta_{z} \\ cos(\phi-g \theta) + \omega_{0} \beta_{\rho}  \big\}\hat{\theta}
	\end{split}
\end{equation}
\begin{equation}\label{chap5_eqn19}
	\frac{d p_{z}}{d t}= a_{0} \beta_{\rho}  sin(\phi-g \theta) - a_{0} g  \beta_{\theta} cos(\phi-g \theta)
\end{equation}
Here $\vec{p}_{\perp} =p_{\rho} \hat{\rho}+ p_{\theta} \hat{\theta}  (=p_{x} \hat{x} + p_{y} \hat{y})$. The $p_{\rho}$, $p_{\theta}$ and $p_{z}$ are the momentum of the particle along $\rho$, $\theta$ and $z$ directions, respectively; and $\beta_{\rho}$, $\beta_{\theta}$ and $\beta_{z}$ are the corresponding velocities. Equation (\ref{chap5_eqn17}) can be further simplified by adding and subtracting the term $-g\left\{a_{0} sin(\phi-g \theta) \dot{\theta} \hat{\rho} + a_{0} g cos(\phi-g\theta) \dot{\theta} \hat{\theta} \right\}$ and then rearranging it, gives

 \begin{equation*}
 	\begin{split}
 		\frac{d \vec{p}_{\perp}}{d t}=  \Bigg\{a_{0}  sin(\phi-g \theta) \frac{d\{t-z-g\theta\}}{dt}\hat{\rho} -a_{0}  cos(\phi-g\theta) \dot{\theta} \hat{\theta} \Bigg\}  - \Bigg\{ a_{0} g  cos(\phi-g \theta)\\ \frac{d\{t-z-g\theta\}}{dt} \hat{\theta} - a_{0} g sin(\phi-g \theta) \dot{\theta} \hat{\rho} \Bigg\}  +  \omega_{0} \left(\beta_{\theta} \hat{\rho}- \beta_{\rho}\hat{\theta}\right)
 	\end{split}
 \end{equation*}
Let's use $\beta_{ \rho }=\dot{ \rho }$, $\beta_{\theta}= \rho \dot{\theta}$, $d\hat{\rho}/dt=\dot{\theta}\hat{\theta}$ and $d\hat{\theta}/dt =-\dot{\theta} \hat{\rho}$, we get
\begin{equation*}
	\begin{split}
			\frac{d \vec{p}_{\perp}}{d t} =  -\Bigg\{a_{0}\frac{d cos(\phi-g \theta)}{dt}\hat{\rho} + a_{0}  cos(\phi-g\theta) \frac {d\hat{\rho}}{dt} \Bigg\}  - \Bigg\{ a_{0} g  \frac{d sin(\phi-g \theta)}{dt} \hat{\theta} + a_{0} g sin(\phi-g \theta)\\ \frac{d\hat{\theta}}{dt} \Bigg\} -  \omega_{0} \left(\rho \frac{d \hat{\theta}}{dt}+ \frac{d {\rho}}{dt}\hat{\theta}\right)
	\end{split} 
\end{equation*}
	or
\begin{equation*}
	\begin{split}
	\frac{d \vec{p}_{\perp}}{d t}	=  -\frac{d}{dt}\left\{a_{0}cos(\phi-g \theta)\hat{\rho} + a_{0} g sin(\phi-g \theta) \hat{\theta} + \omega_{0} \rho \hat{\theta}\right\}
	\end{split} 
\end{equation*}
Integrating and using equation (\ref{chap5_eqn16}), we get 
\begin{equation}\label{perp}
	\begin{split}
		\vec{p}_{\perp} + \vec{A} + \omega_{0} \rho \hat{\theta} = \vec{P}_{\perp 0}
	\end{split} 
\end{equation}
Here $\vec{P}_{\perp 0} (=P_{x0}\hat{x} + P_{y0} \hat{y} = \{(P_{x0} cos(\theta)+ P_{y0} sin(\theta)) \hat{\rho} + \left(- P_{x0} sin(\theta)+ P_{y0} cos(\theta)\right) \hat{\theta} \}$) is constant of motion. 
%
%
The components of equation (\ref{perp}) along $\hat{\rho}$ and $\hat{\theta}$ respectively are 
\begin{equation}\label{chap5_eqn20}
	\gamma \beta_{\rho}=P_{x0} cos(\theta)+ P_{y0} sin(\theta)- a_{0} cos(\phi-g \theta)
\end{equation}
\begin{equation}\label{chap5_eqn21}
	\gamma \beta_{\theta}=- P_{x0} sin(\theta)+ P_{y0} cos(\theta)- a_{0} g sin(\phi-g \theta) -\omega_{0} \rho
\end{equation}
\subsection{Solution of the equation of motion as a function of $\theta$ and $\rho$}
%

Let's consider the simplest case where we choose zero initial transverse canonical momentum. Now equations (\ref{chap5_eqn20}) and (\ref{chap5_eqn21}) become 
\begin{equation}\label{chap5_eqn22}
	\gamma \frac{d \rho}{d t}= - a_{0} cos(\phi-g \theta)
\end{equation}
\begin{equation}\label{chap5_eqn23}
	\gamma \rho \frac{d \theta}{d t}= - a_{0} g sin(\phi-g \theta) -\omega_{0} \rho
\end{equation}

In the resonance condition $(1-v_{z})={-g \omega_{0}}/{\gamma}$, choice of $\dot{\theta}={-\omega_{0}}/{\gamma}$ satisfies $sin(\phi-g \theta)= 0$ ($cos(\phi-g \theta)= (-1)^{n}$, here $n$ is a positive integer).
For resonant case  $sin(\phi-g \theta)$ is zero in general (for $\vec{P}_{\perp 0}=0$),  as shown in Appendix-(\ref{app3}).

 Using these conditions in equations (\ref{chap5_eqn22}) and (\ref{chap5_eqn23}) leads to the expression
\begin{equation*}
	\rho=(-1)^{n} \frac{a_{0}}{\omega_{0}} (\theta-\theta_{0})
\end{equation*}

The equation $\dot{\theta}={-\omega_{0}}/{\gamma}$ implies that $\theta$ decreases monotonically for positively charged particles and increases monotonically for negatively charged particles. Therefore, the value of $n$ is selected to ensure that $\rho$ must increase monotonically (see section (\ref{results})). Thus for positively charged particles, $n$ is odd, while for negatively charged particles, $n$ is even. This results in the behavior of $(-1)^{n}$ being analogous to that of $g$. Thus, the expression for $\rho$ can be rewritten as:   
\begin{equation}\label{chap5_eqn24}
	\rho= g \frac{a_{0}}{\omega_{0}} (\theta-\theta_{0})
\end{equation}
The trajectory shown in equation (\ref{chap5_eqn24}) is an Archimedean spiral, which shows that  a monotonic increment in the radius with $(\theta-\theta_{0})$ (here, $\theta_{0}$ is the value of $\theta$ at $t=0$). Using condition $sin(\phi-g \theta)=0$ in the equation (\ref{chap5_eqn19}), the longitudinal momenta and z-coordinate of the particle can be express in terms of $\theta$ as
\begin{equation}\label{chap5_eqn25}
p_{z} =	 {p_{z 0}} + \frac{ a_{0}^{2}}{2\alpha  } {(\theta-\theta_{0})}^2
\end{equation}
\begin{equation}\label{chap5_eqn26}
	z  =  \frac{g a_{0}^{2}  }{6 {\omega_{0}}^{2} } {(\theta-\theta_{0})}^3 - \frac{p_{z0}}{ \omega_{0}}(\theta-\theta_{0})
\end{equation}
Using equation (\ref{chap5_eqn25}) in $\gamma-p_{z}=\gamma_{0}-p_{z0}$, we get
\begin{equation}\label{chap5_eqn27}
	\gamma  = \gamma_{0} + \frac{ a_{0}^{2}}{2\alpha} {(\theta-\theta_{0})}^2
\end{equation}
We employed the relation $g\omega_{0}=-\alpha$ with $\alpha$ being a positive value representing $|g\omega_{0}|$ (as discussed in the section (\ref{sec_2}), in order to meet the resonance condition for circularly polarized EM wave $g$ and $\omega_{0}$ must have opposite signs). Equations from (\ref{chap5_eqn24})-(\ref{chap5_eqn27}) represent the solution of charged particle dynamics in terms of $(\theta-\theta_{0})$. The expressions in terms of $\rho$ can also be obtained by using the relation $(\theta-\theta_{0})= (g {\omega_{0}}/{a_{0}})\rho$ in above equations. 
\subsection{Solution of equation of motion as a function of $z$}
Equation (\ref{chap5_eqn26}) can be re-written as
\begin{equation}\label{chap5_eqn26_2}
 \frac{g a_{0}^{2}  }{6 {\omega_{0}}^{2} } {(\theta-\theta_{0})}^{3} - \frac{p_{z0}}{ \omega_{0}}(\theta-\theta_{0}) -z = 0
\end{equation}
The discriminant for equation (\ref{chap5_eqn28}) may be written by comparing it with the depressed cubic equation \cite{J_R_Young,loney1904elements,abramowitz1988handbook,gradshteyn2014table} $a(\theta-\theta_{0})^{3} + b (\theta-\theta_{0})+ c =0$ as 
\begin{equation}\label{Dz}
	D_{z} = -\frac{2a_{0}^{2}}{3 \omega_{0}^{4} \alpha}\Bigg(p_{z0}^{3}+\frac{9\alpha a_{0}^{2} z^{2}}{8}\Bigg)
\end{equation}
If $p_{z0}\ge 0$, then the discriminant $D_z$ is negative, meaning that the equation (\ref{chap5_eqn26_2}) can only have a single real root, which is expressed as:

\begin{equation}\label{solution_in_z}
(\theta - \theta_{0} ) = -2g\frac{{\omega_{0}}^{2}}{ a_{0}^{2} }	\Bigg(C_{z}+\frac{D_{z0}}{C_{z}}\Bigg)
\end{equation}
 Here $C_{z}=\left\{\left(-D_{z1} + \sqrt{D_{z1}^{2}-4D_{z0}^{3} }\right)/{2}\right\}^{1/3}$, $D_{z0} = - {a_{0}^{2} p_{z0} }/{ (2\omega_{0}^{2} \alpha)}  $  and  $D_{z1}= -  \{{3 a_{0}^{4}  }/{(4 \omega_{0}^{4})}\}z $.
 Using equation $(\ref{solution_in_z})$ into the aforementioned equations written in terms of $(\theta - \theta_{0})$ for position, momentum, and energy can be expressed in terms of $z$-coordinate (laboratory position). 
 It should be noted that the expressions derived from the equations hold true for initial longitudinal momentum values $p_{z0} \geq 0$. For $p_{z0}<0$ the discriminant $D_{z}>0$ is positive, results in three real roots of the equation (\ref{chap5_eqn26_2}). Thus, for a given value of $z$, the value of $(\theta-\theta_{0})$, momentum, and energy may be multivalued. In this scenario one real root is given by equation (\ref{solution_in_z}) (say $(\theta-\theta_{0})_{1}$) and other two roots can be expressed as
 \begin{equation}\label{solution_in_z_back}
 	\left(\theta - \theta_{0} \right) = -2g\frac{{\omega_{0}}^{2}}{ a_{0}^{2} }	\Bigg\{\xi_{i} C_{z}+\frac{D_{z0}}{\xi_{i} C_{z}}\Bigg\}
 \end{equation} 
Here, index $i$ ranges from $1$ to $2$, while $\xi_{1}$ and $\xi_{2}$ represent two distinct complex roots of unity, defined as $\left(-1+\iota \sqrt{3}\right)/2$ and $\left(-1-\iota \sqrt{3}\right)/2$, respectively. It can be seen from the expression of discriminant (equation (\ref{Dz})) that as dynamics evolves and the value of $z$ changes from zero, the second term of the equation (\ref{Dz}) starts increasing. 
As the magnitude of the term $9\alpha a_{0}^{2} z^{2}/8$ surpasses that of $|p_{z0}^{3}|$, it leads to the condition $D_{z} \ge 0$, subsequently causing the equation to exhibit a real root.
Thus multiple roots can only exists in the domain  $ |z| \le (2|p_{z0}|/3a_{0}) \sqrt{2|p_{z0}|/\alpha}$. The physically consistent roots may be retain from the above expression in the following way. The equation $\dot{\theta} (= -\omega_{0}/\gamma)=g(1-v_{z})$  implies that the evolution of $(\theta - \theta_0)$ over time exhibits a monotonic pattern, which consequently results in a monotonic increase in the variable $p_z$ (as shown in equation (\ref{chap5_eqn25})). This behavior of $p_{z}$ ensures that for the choice of $p_{z0}<0$ dynamics of the charged particle must reaches to a point after which charged particle reflects back and starts moving along the wave propagation direction. The value of $(\theta - \theta_0)$ at the reflection point may be calculated by putting $p_{z}=0$ in equation (\ref{chap5_eqn25}), that gives $(\theta - \theta_0)_{ref} =-g{\sqrt{2 \alpha |p_{z0}|} }/{a_{0}}$ or $( \rho_{ref} = {\sqrt{2 a_{0} |p_{z0}|} }/{\alpha} )$. Substituting $(\theta - \theta_0)_{ref}$ in equation (\ref{chap5_eqn26}) gives  
\begin{equation}\label{reflection_point}
z_{ref} =\frac{ -2 |p_{z0}| }{3 a_{0}}\sqrt{\frac{2 |p_{z0}|}{\alpha}} 
\end{equation} 
Here, the parameters $(\theta-\theta_{0})_{ref}$, $\rho_{ref}$, and $z_{ref}$ correspond to the angular ($\theta$), radial ($\rho$), and vertical ($z$) coordinates evaluated at the point of reflection, respectively. 
It is important to note that the constant of motion, $\Delta = (\gamma - p_{z})$, implies that the resonant energy gain occurs through a corresponding monotonically increasing momentum along the wave propagation direction (longitudinal momentum). Consequently, for any choice of $p_{z0} < 0$ (where $z = 0$ at $t = 0$), the reflection point exists exclusively in the region where $z < 0$. Therefore, equation (\ref{chap5_eqn26_2}) has a unique real root within the domain of $z > 0$.
Using this and equation (\ref{reflection_point}) in the context for domain of multiple roots, the interval where multiple roots exist can be redefined as $-|z_{ref}| < z < 0$. Thus, any root which exists for $z<z_{ref}$ must not be physically consistent. That can be investigated by substituting $z = z_{ref}$ into equations (\ref{solution_in_z}) and (\ref{solution_in_z_back}), and subsequently comparing the resulting $(\theta-\theta_{0})$ values with $\theta_{ref}$ (this may give a double point at which two roots must meet and one root should has value different than $\theta_{ref}$). By doing so it can be seen that root given in equation (\ref{solution_in_z}) and root corresponds to $\xi_{2}$ (say $(\theta-\theta_{0})_{3}$) in equation (\ref{solution_in_z_back})  result $(\theta-\theta_{0}) = \theta_{ref}$. Whereas, root corresponds to $\xi_{1}$ (say $(\theta-\theta_{0})_{2}$) in equation (\ref{solution_in_z_back}) gives  $(\theta-\theta_{0}) =2 \theta_{ref}$. Thus, $(\theta-\theta_{0})_{1}$ and $(\theta-\theta_{0})_{3}$ are the only two roots of equation (\ref{chap5_eqn26_2}) that describes the physical picture of the charge particle dynamics. 
 Thus, the solution to equation (\ref{chap5_eqn26_2}), which completely describes the dynamics of the charged particle, can be expressed as:

        \begin{equation}\label{theta_vs_z} 
        	(\theta-\theta_{0})=
        	\begin{cases}
        		-2g\frac{{\omega_{0}}^{2}}{ a_{0}^{2} }	\Bigg\{ C_{z}+\frac{D_{z0}}{ C_{z}}\Bigg\} ,& \text{for } -|z_{ref}| \geq z\leq \infty\\	
        		
        		-g\frac{{\omega_{0}}^{2}}{ a_{0}^{2} }	\Bigg\{\left(-1+\iota \sqrt{3}\right) C_{z}+\frac{4 D_{z0}}{\left(-1+\iota \sqrt{3}\right) C_{z}}\Bigg\},  & \text{for } -|z_{ref}| \geq z\leq 0
        	\end{cases}
         \end{equation}                                            
Substituting equation (\ref{theta_vs_z}) in equations (\ref{chap5_eqn24}), (\ref{chap5_eqn25}), (\ref{chap5_eqn26}) and (\ref{chap5_eqn27}) gives position, momentum and energy of the charged particle as a function of $z$. 

\begin{figure}
	\includegraphics[width= 15 cm, height=8 cm]{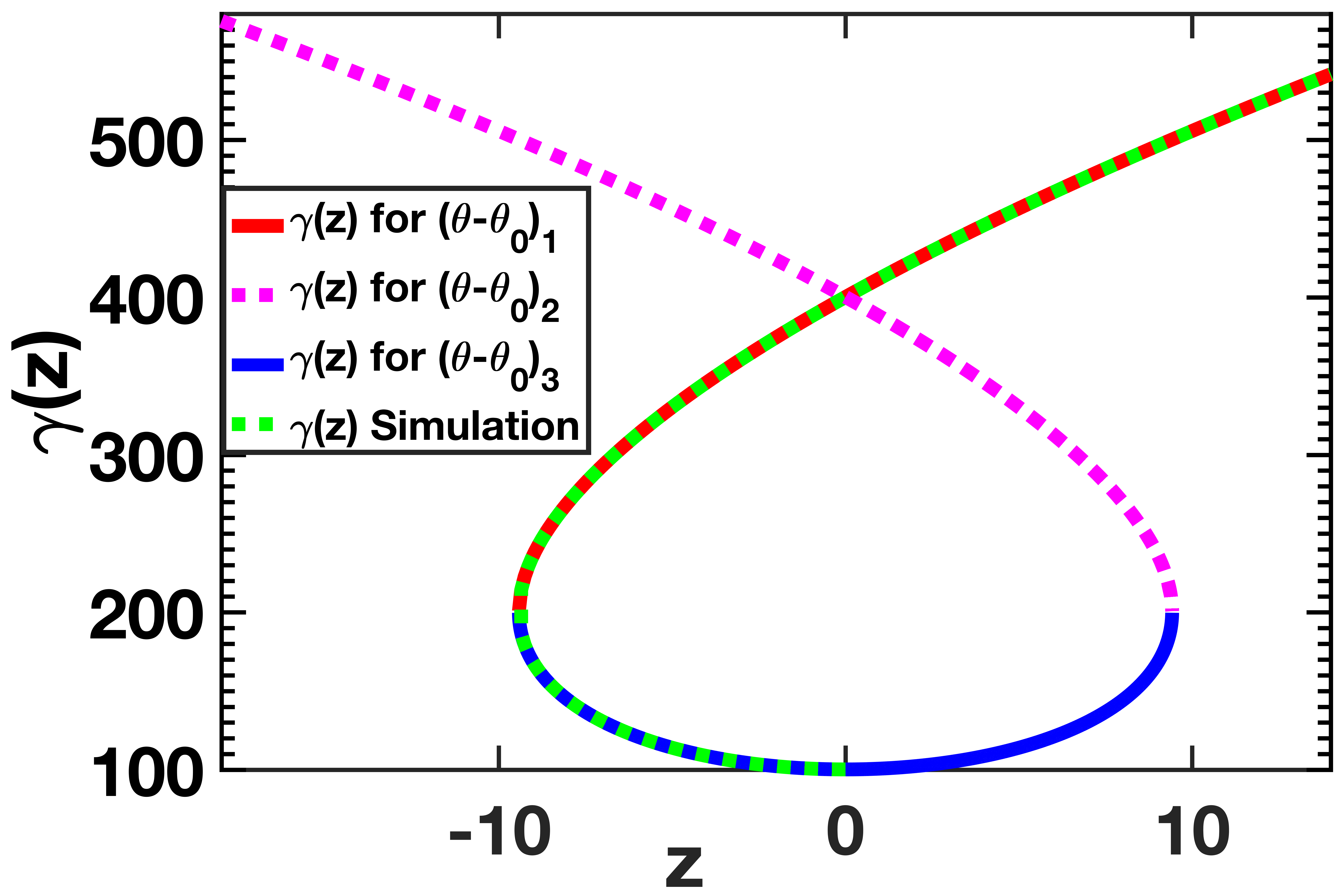}
	\caption{The figure represents energy of positive charged particle in relation to  z-coordinate of its position. These graphs are plotted for a charged particle with initial momenta $p_{x0}= -10/\sqrt{2}$, $p_{y0}= 0$, and $p_{z0}= -100$, resonantly ($\Omega_{0}=1$) interacting with a circularly polarized ($\delta=1/\sqrt{2}$) EM wave characterized by $\phi_{0}=0$, $a_{0}=10$, and $g = -1$. The red, magenta, and blue curves respectively represent the energy of the charged particle correspond to roots $(\theta - \theta_{0})_{1}$, $(\theta - \theta_{0})_{2}$, and $(\theta - \theta_{0})_{3}$.  Whereas the green curve represents energy of the particle obtained using test particle simulation.}
	\centering
	\label{gamma_vs_z_ref}
\end{figure}

 For the further verification of physical arguments and our analytical results (equation (\ref{theta_vs_z})), we shall compare these results with the results obtained from the numerical code.
The energy of the charged particle for $p_{z0}<0$ is shown in the figure \ref{gamma_vs_z_ref}. This figure may be divided into two parts:
 Firstly motion of the charged particle described by part of the blue curve lies in the domain $z>0$ and magenta curve. These curves represent gain in energy along with gain in momentum of the charged particle opposite to the wave propagation direction. This leads to monotonic increase in the $\Delta (=\gamma-p_{z})$ and hence represents the unphysical part of the curve. Secondly part of the blue curve lies in the domain $z<0$ and red curve. These curves represent gain in energy of the charged particle with the gain in momentum along the wave propagation direction. This is the only possible way in which $\Delta (=\gamma-p_{z})$ is a constant of motion and physically consistent part of the curve. The points of contact for red and blue curves in the domains of $z > 0$ shows the  reflection point and satisfies $\theta=\theta_{ref}(\approx 28.302)$ at $z=z_{ref}(\approx -9.422)$. Whereas the magenta curve corresponds to $\theta=2\theta_{ref}(\approx 56.604)$ at $z=z_{ref}(\approx -9.422)$. It further shows that the solution described by $(\theta-\theta_{0})_{2}$ is unphysical and physically correct solution is described by the equation (\ref{theta_vs_z}). It has also been verified by compering the energy obtained by using test particle simulation (shown by green curve).        



  
\subsection{Solution of equation of motion as a function of lab time $t$}
Now using the energy equation (\ref{chap5_eqn27}) in the equation $\dot{\theta}={-\omega_{0}}/{\gamma}$ gives
\begin{equation*}
	\gamma d\theta =\left\{\gamma_{0} + \frac{ a_{0}^{2}}{2\alpha} {(\theta-\theta_{0})}^2\right\}d\theta = -\omega_{0} dt
\end{equation*}
On integrating, one can get
\begin{equation*}
	\gamma_{0}(\theta-\theta_{0}) + \frac{ a_{0}^{2}}{6\alpha} {(\theta-\theta_{0})}^3 = -\omega_{0} t
\end{equation*}
Using $g^{2} = 1$ in above expression, we get
\begin{equation}\label{chap5_eqn28}
	\frac{a_{0}^{2}}{6 \alpha }(\theta-\theta_{0})^{3}+\gamma_{0} (\theta-\theta_{0}) -g \alpha t =0 
\end{equation}
 The discriminant for equation (\ref{chap5_eqn28}) may be written by comparing it with the depressed cubic equation \cite{J_R_Young,loney1904elements,abramowitz1988handbook,gradshteyn2014table} $a(\theta-\theta_{0})^{3} + b (\theta-\theta_{0})+ c =0$ as  
\begin{equation}\label{chap5_eqn30}
	D_{t} = -4 a b^{3}-27 a^{2} c^{2} = - \frac{a_{0}^{2} }{6 \alpha }\Bigg(4 {\gamma_{0}}^{3}+ 9 \frac{a_{0}^{2}  \alpha }{2} t^{2}\Bigg)<0
\end{equation}
The discriminant $D_{t}$ turns out to be negative. Hence equation (\ref{chap5_eqn28}) has only one real root, which is given by
\begin{equation}\label{chap5_eqn31}
	(\theta-\theta_{0}) = -\frac{2 \alpha}{a_{0}^{2}}\Bigg(C_{t}+\frac{D_{t0}}{C_{t}}\Bigg)
\end{equation}
Here $C_{t}=\{(D_{t1} + \sqrt{D_{t1}^{2}-4D_{t0}^{3} })/{2}\}^{1/3}$;
and $D_{t0} = - {a_{0}^{2} \gamma_{0}}/{ (2 \alpha)}  $  and  $D_{t1}= -  \{{3g a_{0}^{4}  }/{(4 \alpha)}\}t $. 	
Since $D_{t0}$ is a negative number, thus $C_{t}$ can not be an imaginary number. 
Substituting equation $(\ref{chap5_eqn31})$ in the equations (\ref{chap5_eqn24})-(\ref{chap5_eqn27}) yields a solution for the equation of motion in terms of laboratory time. This solution applies to all possible choices of initial longitudinal momentum. 
\vspace{-1 cc}
\section{Summary and conclusions}\label{Summary} In the present work interaction of a charged particle with elliptically polarized EM wave and an axial magnetic field has been analytically and numerically studied. In this regard, we consider the dynamics of a charged particle with arbitrary initial momentum placed at the origin ($\vec{r}_{0} = 0$) interacting with an arbitrary initial phase of a polarized transverse EM wave. This choice does not lose the generality of the problem. Next, the dynamics of charged particle, with respect to the phase of an EM wave, is expressed as that of a forced driven harmonic oscillator. These dynamical equations implies that for the cyclotron frequency of the particle is equal to Doppler shifted frequency of the wave, resonance occurs, and the particle gains unbounded energy from the wave \cite{kolomenskii1962autoresonance,kolomenskii1963self, DavydovskyVa,Ondarza-Rovira}. 
%
Furthermore, it has been demonstrated that an elliptically polarized EM wave can be represented as the combination of two circularly polarized EM waves with different amplitudes and opposite helicity. To achieve resonance in an elliptically polarized EM wave, it is necessary for the cyclotron rotation of the charged particle to align both in direction and rate with the rotation of the electric field vector of one of its constituent circularly polarized EM waves. This leads to the dependency of the resonant condition on helicity and energy gain on choice of $\delta$.    
In this regard, it has been shown that for the choice of $g \Omega_{0}=-1$ the net energy gain by the charged particle is maximum for the circularly polarized EM wave and minimum for the linearly polarized EM wave. On the other hand, for the choice of $g \Omega_{0}=1$ no net energy gain in case of circularly polarized EM wave and maximum energy gain for the linearly polarized EM wave. The net energy gain described for arbitrary polarization (except for linearly polarization) is always larger for $g \Omega = -1$ case compared to $g \Omega = 1$ case. Whereas, the net energy gain turns out to be same for linearly polarized EM wave for both the cases ($g \Omega = -1$ or $g \Omega = 1$). Furthermore, the net energy gain for arbitrary polarization is represented in terms of energy gain in the circularly polarized EM wave using the relationship given in equation (\ref{dif_delta}).
This analysis shows that study of cyclotron auto-resonance in a circularly polarized EM wave can serve a foundation for understanding of phase locking and net energy gain 
in the context of particle acceleration through cyclotron auto-resonance in an elliptically polarized EM wave. Thus, in the subsequent section, 
we present an explicit solution to the governing equation in terms of particle position or laboratory time for the case of cyclotron auto-resonance in a circularly polarized EM wave. 
These explicit expressions are useful in the study of microwave generation, plasma heating, particle acceleration and cosmic ray generation, etc.  
\section{APPENDIX-1}\label{app1}
\begin{equation}
\begin{split}
x(\phi)= -\frac{1}{\Delta  m \omega  \Omega _0 (\Omega _{0}^2-1)}(P_{y0} + \Delta  m w_{1} \omega  \Omega _0 \sin (\phi_{0}) \sin (\Omega _0 \phi ) \sin (\Omega _0 \phi _0)+\Delta  m w_{1} \omega  \Omega _0^2\\
 \cos (\phi_{0}) \sin (\Omega _0 \phi ) \cos (\Omega _0 \phi _0)-\Delta  m w_{1} \omega  \Omega _0^2 \cos (\phi_{0})\sin (\Omega _0 \phi _0)\cos (\Omega _0 \phi )+\Delta  m w_{1} \omega  \Omega _0 \\
 \sin (\phi_{0}) \cos (\Omega _0 \phi ) \cos (\Omega _0 \phi _0)
+\Delta  m \omega  \Omega _0 \sin (\phi ) (w_{2} \Omega _0-w_{1})-\Delta  m w_{2} \omega \Omega _0^2 \sin (\phi_{0}) \\
\sin (\Omega _0 \phi )\sin (\Omega _0 \phi _0)-\Delta  m w_{2} \omega  \Omega _0^2 \sin (\phi_{0}) \cos (\Omega _0 \phi ) \cos (\Omega _0 \phi _0)-\Delta  m w_{2} \omega  \Omega _0 \cos (\phi_{0})] \\
\sin (\Omega _0 \phi ) \cos (\Omega _0 \phi _0)+\Delta  m w_{2} \omega  \Omega _0 \cos (\phi_{0}) \sin (\Omega _0 \phi _0)\cos (\Omega _0 \phi )-\Omega _0^2 P_{x0} \sin (\Omega _0 (\phi -\phi_{0}))\\
+P_{x0} \sin (\Omega _0 (\phi -\phi_{0}))
-\Omega _0^2 P_{y0}+(\Omega _0^2-1) P_{y0} \cos (\Omega _0 (\phi -\phi_{0})) )
\end{split}
\end{equation}
\begin{equation}
\begin{split}
y(\phi)= \frac{1}{\Delta  m \omega  (\Omega _0-1) \Omega _0 (\Omega _0+1)} (-\Delta  m w_{1} \omega  \Omega _0^2 \cos (\phi _0) \cos (\Omega _0 \phi ) \cos (\Omega _0 \phi _0)
-\Delta  m w_{1} \omega \\
 \Omega _0^2 \cos (\phi _0)
\sin (\Omega _0 \phi )  \sin (\Omega _0 \phi _0)+\Delta  m w_{1} \omega  \Omega _0 \sin (\phi _0) \sin (\Omega _0 \phi ) \cos (\Omega _0 \phi _0)-\Delta  m w_{1} \omega  \Omega _0 \\
\sin (\phi _0) \sin (\Omega _0 \phi _0)
 \cos (\Omega _0 \phi )+\Delta  m \omega  \Omega _0 \cos (\phi ) (w_{1} \Omega _0-w_{2})+\Delta  m w_{2} \omega  \Omega _0 \cos (\phi _0) \cos (\Omega _0 \phi ) \\
 \cos (\Omega _0 \phi _0)-\Delta  m w_{2} \omega 
  \Omega _0^2 \sin (\phi _0) \sin (\Omega _0 \phi ) \cos (\Omega _0 \phi _0)+\Delta  m w_{2} \omega  \Omega _0^2 \sin (\phi _0) \sin (\Omega _0 \phi _0) 
  \cos (\Omega _0 \phi )\\
  +\Delta  m w_{2} \omega  \Omega _0 \cos (\phi _0)
   \sin (\Omega _0 \phi ) \sin (\Omega _0 \phi _0)-\Omega _0^2 P_{{x0}}
  +(\Omega _0^2-1) P_{{x0}}\cos (\Omega _0 (\phi -\phi _0))+P_{{x0}}+\Omega _0^2 \\
  P_{{y0}} \sin (\Omega _0 (\phi -\phi _0))-P_{{y0}} \sin (\Omega _0 (\phi -\phi _0)))
\end{split}
\end{equation}
\begin{equation*}
\begin{split}
z(\phi)= \frac{1}{4 m \omega  (\Omega _0^2-1){}^2}(4 m \Delta  \omega  \Omega _0^4-8 m \Delta  \omega  \Omega _0^2+4 m \Delta  \omega +k (4 p_{z0} \Omega_0^4-8 p_{z0} w_{0}
 \Omega_{0}^{2}+4 p_{z0})\\
 +k^2 (-4 m w_{1} w_{2} \Delta  \omega  \Omega _0^3 +4 m w_{1} w_{2} \Delta  \omega  \cos (\Omega _{0} \phi) \cos (\Omega _0 \phi _0 ) \sin (\phi ) \sin (\phi _0) \Omega _0^3+4 m w_{2}^2 \\ 
  \Delta  \omega  \cos (\phi )\cos (\Omega _0 \phi _0) \sin (\phi _0) \sin (\Omega _0 \phi) \Omega _0^3 
 -4 m w_{2}^2 \Delta  \omega  \cos (\phi ) \cos (\Omega _0 \phi) sin (\phi _0)\\
  \sin (\Omega _0 \phi _0) \Omega _0^3+4 m w_{1} w_{2} \Delta  \omega  \sin (\phi ) \sin (\phi _0) \sin (\Omega _0 \phi) \sin (\Omega _0 \phi _0)\Omega _0^3  -4 w_{2} \cos (\phi )\\
  \cos (\Omega _0 \phi)\cos (\Omega _0 \phi _0)  P_{{x0}}\Omega _0^3+4 w_{1}\cos (\Omega _0 \phi _0) \sin (\phi ) \sin (\Omega _0 \phi) P_{{x0}} \Omega _0^3-4 w_{1} \cos (\Omega _0 \phi)\\  
  \sin (\phi ) \sin (\Omega _0 \phi _0) P_{{x0}} \Omega _0^3-4 w_{2} \cos (\phi ) \sin (\Omega _0 \phi) \sin (\Omega _0 \phi _0)P_{{x0}} \Omega _0^3-4 w_{1}\cos (\Omega _0 \phi)\\
\cos (\Omega _0 \phi _0) \sin (\phi ) P_{{y0}} \Omega _0^3 +4 w_{1}  \sin (\phi _0) P_{{y0}} \Omega _0^3-4 w_{2} \cos (\phi ) \cos (\Omega _0 \phi _0) \sin (\Omega _0 \phi)\\ 
P_{{y0}} \Omega _0^3+4 w_{2} \cos (\phi ) \cos (\Omega _0 \phi)  \sin (\Omega _0 \phi _0)P_{{y0}} \Omega _0^3-4 w_{1} \sin (\phi ) \sin (\Omega _0 \phi) \sin (\Omega _0 \phi _0) \\
P_{{y0}} \Omega _0^3+4 m w_{1}^2 \Delta  \omega \Omega _0^2+4 m w_{2}^2 \Delta  \omega \Omega _0^2+m w_{1}^2 \Delta  \omega  \cos (2 \phi _0) \Omega _0^2-m w_{2}^2 
\Delta  \omega  \cos (2 \phi _0)\\
 \Omega _0^2 -4 m w_{1}^2 \Delta  \omega  \cos (\Omega _0 \phi) \cos (\Omega _0 \phi _0) \sin (\phi ) \sin (\phi _0)  \Omega _0^2 -4 m w_{2}^2 \Delta  \omega \cos (\Omega _0 \phi)\\
 \cos (\Omega _0 \phi _0)\sin (\phi ) \sin (\phi _0)\Omega _0^2-8 m w_{1} w_{2} \Delta  \omega  \cos (\phi ) \cos (\Omega _0 \phi _0) \sin (\phi _0)\sin (\Omega _0 \phi) \Omega _0^2\\
 +8 m w_{1} w_{2} \Delta  \omega \cos (\phi ) \cos (\Omega _0 \phi) \sin (\phi _0) 
  \sin (\Omega _0 \phi _0) \Omega _0^2-4 m w_{1}^2 \Delta \omega  \sin (\phi )\sin (\phi _0)\\
\sin (\Omega _0 \phi) \sin (\Omega _0 \phi _0) \Omega _0^2-4 m w_{2}^2 \Delta  \omega  \sin (\phi )
  \sin (\phi _0)  \sin (\Omega _0 \phi) \sin (\Omega _0 \phi _0) \Omega _0^2 
    +4 w_{1} \\
       \cos (\phi ) \cos (\Omega _0 \phi _0) \cos (\Omega _0 \phi _0) P_{{x0}} \Omega _0^2-4 w_{2} \cos (\Omega _0 \phi _0)  \sin (\phi ) \sin (\Omega _0 \phi) P_{{x0}} \Omega _0^2 +4 w_{2}\\ 
    \cos (\Omega _0 \phi) \sin (\phi) \sin (\Omega _0 \phi _0) P_{{x0}} \Omega _0^2+4 w_{1} \cos (\phi)  
    \sin (\Omega _0 \phi) \sin (\Omega _0 \phi _0) P_{{x0}} \Omega _0^2+4 w_{2} \\
    \cos (\Omega _0 \phi) \cos (\Omega _0 \phi _0)
    \sin (\phi ) P_{{y0}} \Omega _0^2-4 w_{2} \sin (\phi _0) P_{{y0}} \Omega _0^2+4 w_{1} \cos (\phi ) \cos (\Omega _0 \phi _0)\\
    \sin (\Omega _0 \phi) P_{{y0}} \Omega _0^2-4 w_{1} \cos (\phi ) \cos (\Omega _0 \phi) \sin (\Omega _0 \phi _0) P_{{y0}} \Omega _0^2 +4 w_{2} \sin (\phi ) \sin (\Omega _0 \phi)\\
\end{split}
\end{equation*}  
\begin{equation*}
	\begin{split}     
    \sin (\Omega _0 \phi _0) P_{{y0}} \Omega _0^2-4 m w_{1} w_{2} \Delta  \omega  \Omega _0 
    +4 m w_{1} w_{2} \Delta  \omega  \cos (\Omega _0 \phi) \cos (\Omega _0 \phi _0) \sin (\phi )\\
 sin (\phi _0)\Omega _0 + 4 m w_{1}^2 \Delta  \omega  \cos (\phi ) \cos (\Omega _0 \phi _0) \sin (\phi _0) \sin (\Omega _0 \phi) \Omega _0-4 m w_{1}^2 \Delta  \omega  \cos (\phi )\\ 
 \cos (\Omega _0 \phi) \sin (\phi _0) \sin (\Omega _0 \phi _0) \Omega _0 +4 m w_{1} w_{2} \Delta  \omega  \sin (\phi ) \sin (\phi _0) \sin (\Omega _0 \phi) \sin (\Omega _0 \phi _0) \Omega _0 \\
  +4 w_{2} \cos (\phi ) \cos (\Omega _0 \phi) \cos (\Omega _0 \phi _0) P_{{x0}} \Omega _0-4 w_{1} \cos (\Omega _0 \phi _0) \sin (\phi ) \sin (\Omega _0 \phi) P_{{x0}}\Omega _0+4 \\
  w_{1} \cos (\Omega _0 \phi) \sin (\phi ) \sin (\Omega _0 \phi _0) P_{{x0}} \Omega _0+4 w_{2} \cos (\phi ) \sin (\Omega _0 \phi) \sin (\Omega _0 \phi _0) P_{{x0}} \Omega _0+4 w_{1}\\       
\end{split}
\end{equation*}
\begin{equation}
\begin{split}
\begin{aligned}
	\cos (\Omega _0 \phi) \cos (\Omega _0 \phi _0) \sin (\phi ) P_{{y0}} \Omega _0 -4 w_{1} \sin (\phi _0) P_{{y0}} \Omega _0+4 w_{2}  \cos (\phi ) \cos (\Omega _0 \phi _0) \\
	 \sin (\Omega _0 \phi) P_{{y0}} \Omega _0-4 w_{2} \cos (\phi ) \cos (\Omega _0 \phi) \sin (\Omega _0 \phi _0) P_{{y0}} \Omega _0 + 4 w_{1} \sin (\phi )\sin (\Omega _0 \phi) \\
	\sin (\Omega _0 \phi _0) P_{{y0}} \Omega _0-m w_{1}^2 \Delta  \omega  \cos (2 \phi _0)
	+m w_{2}^2 \Delta  \omega  \cos (2 \phi _0)-4 w_{1} \cos (\phi ) \cos (\Omega _0 \phi)\\  
	\cos (\Omega _0 \phi _0) P_{{x0}}+4 w_{2} \cos (\Omega _0 \phi _0) \sin (\phi ) \sin (\Omega _0 \phi) P_{{x0}}-4 w_{2} \cos (\Omega _0 \phi) \sin (\phi )\sin (\Omega _0 \phi _0) \\
	P_{{x0}}-4 w_{1}\cos (\phi ) \sin (\Omega _0 \phi) \sin (\Omega _0 \phi _0) P_{{x0}}-4 w_{2} \cos (\Omega _0 \phi) \cos (\Omega _0 \phi _0) \sin (\phi )P_{{y0}}+4 \\
w_{2} \sin (\phi _0) P_{{y0}}-4 w_{1} \cos (\phi ) \cos (\Omega _0 \phi _0) \sin (\Omega _0 \phi) P_{{y0}} +4 w_{1} \cos (\phi ) \cos (\Omega _0 \phi) \sin (\Omega _0 \phi _0) P_{{y0}}\\
-4 w_{2} \sin (\phi ) \sin (\Omega _0 \phi) \sin (\Omega _0 \phi _0) P_{{y0}}-m (w_{1}^2-w_{2}^2) \Delta  \omega  \cos (2 \phi ) (\Omega _0^2-1) +4 \cos (\phi _0) (m w_{1}^2 \Delta \\
 \omega  \cos (\Omega _0 \phi) \sin (\phi )
 \sin (\Omega _0 \phi _0)\Omega _0^3+w_{2} P_{{x0}}\Omega _0^3-2 m w_{1} w_{2} \Delta  \omega  \cos (\Omega _0 \phi) \sin (\phi ) \sin (\Omega _0 \phi _0) \Omega _0^2\\
 -w_{1}  P_{{x0}} \Omega _0^2-m \Delta  \omega  \cos (\Omega _0 \phi _0)
 \sin (\phi ) \sin (\Omega _0 \phi) (w_{2}-w_{1} \Omega _0){}^2 \Omega _0 +m w_{2}^2 \Delta  \omega  \cos (\Omega _0 \phi) \sin (\phi )\\
  \sin (\Omega _0 \phi _0) \Omega _0-w_{2} P_{{x0}} \Omega _0  +m \Delta\omega  \cos (\phi ) \cos ((\phi -\phi _0) \Omega _0) (w_{2} (\Omega _0^2+1) w_{1}-w_{2}^2 \Omega _0-w_{1}^2 \\
 \Omega _0) \Omega _0+w_{1} P_{{x0}})))
\end{aligned}
\end{split}
\end{equation}
\begin{equation}
\begin{split}
p_{x}(\phi) = \frac{P_{x0}}{m \omega \Delta} - w_{1} cos(\phi)+   
\frac{1}{(\Delta  m \omega  (\Omega _0-1) \left(\Omega _0+1\right)}(-\Delta  m w_{1} \omega  \Omega _0^2 \cos (\phi _0) \cos (\Omega _0 \phi ) \\
\cos (\Omega _0 \phi _0)-\Delta  m w_{1} \omega  \Omega _0^2 \cos (\phi _0)
\sin (\Omega _0 \phi )  \sin (\Omega _0 \phi _0)+\Delta  m w_{1} \omega  \Omega _0 \sin (\phi _0) \sin (\Omega _0 \phi )\\
 \cos (\Omega _0 \phi _0)-\Delta  m w_{1} \omega  \Omega _0 \sin (\phi _0) \sin (\Omega _0 \phi _0)\cos (\Omega _0 \phi )+\Delta  m \omega  \Omega _0 \cos (\phi ) (w_{1} \Omega _0-w_{2})\\
 +\Delta  m w_{2} \omega  \Omega _0 \cos (\phi _0) \cos (\Omega _0 \phi ) \cos (\Omega _0 \phi _0)-\Delta  m w_{2} \omega \Omega _0^2 \sin (\phi _0) \sin (\Omega _0 \phi ) \cos (\Omega _0 \phi _0)\\
  +\Delta  m w_{2} \omega  \Omega _0^2 \sin (\phi _0) \sin (\Omega _0 \phi _0) 
  \cos (\Omega _0 \phi )+\Delta  m w_{2} \omega  \Omega _0 \cos (\phi _0)
   \sin (\Omega _0 \phi )\sin (\Omega _0 \phi _0)\\
   -\Omega _0^2 P_{{x0}} +(\Omega _0^2-1) P_{{x0}} \cos (\Omega _0 (\phi -\phi _0))+P_{{x0}}+\Omega _0^2 P_{{y0}} \sin (\Omega _0 (\phi -\phi _0))
  -P_{{y0}} \\
  \sin (\Omega _0 (\phi -\phi _0)))
\end{split}
\end{equation}
\begin{equation}
\begin{split}
p_{y}(\phi) = \frac{P_{y0}}{m \omega \Delta} - w_{2} sin(\phi)-\frac{1}{\Delta  m \omega   (\Omega _{0}^2-1)} ( (P_{y0} + \Delta  m w_{1} \omega  \Omega _0 \sin (\phi_{0}) \sin (\Omega _0 \phi ) \sin (\Omega _0 \phi _0)\\
+\Delta  m w_{1} \omega  \Omega _0^2 \cos (\phi_{0}) \sin (\Omega _0 \phi ) \cos (\Omega _0 \phi _0)-\Delta  m w_{1} \omega  \Omega _0^2 \cos (\phi_{0})\sin (\Omega _0 \phi _0) \cos (\Omega _0 \phi )\\
+\Delta  m w_{1} \omega  \Omega _0 \sin (\phi_{0}) \cos (\Omega _0 \phi ) \cos (\Omega _0 \phi _0)+\Delta  m \omega  \Omega _0 \sin (\phi )(w_{2} \Omega _0-w_{1})-\Delta  m w_{2} \omega \\ 
\Omega _0^2 \sin (\phi_{0}) \sin (\Omega _0 \phi ) \sin (\Omega _0 \phi _0)-\Delta  m w_{2} \omega  \Omega _0^2 \sin (\phi_{0}) \cos (\Omega _0 \phi )
 \cos (\Omega _0 \phi _0)-\Delta  m w_{2} \\
 \omega \Omega _0 \cos (\phi_{0})] \sin (\Omega _0 \phi ) \cos (\Omega _0 \phi _0)+\Delta  m w_{2} \omega  \Omega _0 \cos (\phi_{0}) \sin (\Omega _0 \phi _0)
 \cos (\Omega _0 \phi )-\Omega _0^2 P_{x0}\\
  \sin (\Omega _0 (\phi -\phi_{0}))+P_{x0} \sin (\Omega _0 (\phi -\phi_{0}))-\Omega _0^2 P_{y0}+(\Omega _0^2-1) P_{y0} \cos (\Omega _0 (\phi -\phi_{0})) ))
\end{split}
\end{equation}
\begin{equation*}
\begin{split}
p_{z}(\phi)= p_{z0} + \frac{1}{4 \omega (\Omega _0^2-1)^2}(k^2 \Delta  (4 m w_2^2 \Delta  \omega  \cos (\phi ) \cos (\Omega _0 \phi) \cos (\Omega _0 \phi _0) \sin (\phi _0) \Omega _0^4-4 m {w_1} \\
{w_2} \Delta  \omega  \cos (\Omega _0 \phi _0) \sin (\phi ) \sin (\phi _0) \sin (\Omega _0 \phi) \Omega _0^4+4 m {w_1} {w_2} \Delta  \omega  \cos (\Omega _0 \phi) \sin (\phi ) \sin (\phi _0) \\
\sin (\Omega _0 \phi _0) \Omega _0^4+4 m w_2^2 \Delta  \omega  \cos (\phi ) \sin (\phi _0) \sin (\Omega _0 \phi) \sin (\Omega _0 \phi _0) \Omega _0^4+4 {w_1} \cos (\Omega _0 \phi) \\
\cos (\Omega _0 \phi _0)\sin (\phi ) P_{{x0}} \Omega _0^4+4 {w_2} \cos (\phi ) \cos (\Omega _0 \phi _0) \sin (\Omega _0 \phi) P_{{x0}} \Omega _0^4-4 {w_2} \cos (\phi ) \cos (\Omega _0 \phi)\\ 
\sin (\Omega _0 \phi _0) P_{{x_0}} \Omega _0^4+4 {w_1} \sin (\phi ) \sin (\Omega _0 \phi) \sin (\Omega _0 \phi _0) P_{{x_0}} \Omega _0^4-4 {w_2} \cos (\phi ) \cos (\Omega _0 \phi) \\
\cos (\Omega _0 \phi _0) P_{{y_0}} \Omega _0^4+4 {w_1} \cos (\Omega _0 \phi _0) \sin (\phi ) \sin (\Omega _0 \phi) P_{{y_0}} \Omega _0^4-4 {w_1}\cos (\Omega _0 \phi) \sin (\phi ) \\
\sin (\Omega _0 \phi _0) P_{{y_0}} \Omega _0^4-4 {w_2} \cos (\phi ) \sin (\Omega _0 \phi) \sin (\Omega _0 \phi _0) P_{{y_0}} \Omega _0^4 -4 m {w_1} {w_2} \Delta  \omega  \cos (\phi ) \\
\cos (\Omega _0 \phi) \cos (\Omega _0 \phi _0) \sin (\phi _0) \Omega _0^3+4 m w_1^2 \Delta  \omega  \cos (\Omega _0 \phi _0) \sin (\phi ) \sin (\phi _0) \sin (\Omega _0 \phi) \\
\Omega _0^3-4 m w_1^2 \Delta  \omega  \cos (\Omega _0 \phi) \sin (\phi ) \sin (\phi _0) \sin (\Omega _0 \phi _0) \Omega _0^3 -4 m {w_1} {w_2} \Delta  \omega  \cos (\phi )\\
 \sin (\phi _0) \sin (\Omega _0 \phi) \sin (\Omega _0 \phi _0) \Omega _0^3-4 m {w_2}^2 \Delta  \omega  \cos (\phi ) \cos (\Omega _0 \phi) \cos (\Omega _0 \phi _0) \sin (\phi _0) \Omega _0^2)\\
 +4 m {w_1} {w_2} \Delta  \omega  \cos (\Omega _0 \phi _0) \sin (\phi ) \sin (\phi _0) \sin (\Omega _0 \phi) \Omega _0^2-4 m {w_1} {w_2} \Delta  \omega  \cos (\Omega _0 \phi) \sin (\phi ) \\
 \sin (\phi _0) \sin (\Omega _0 \phi _0) \Omega _0^2-4 m w_2^2 \Delta  \omega  \cos (\phi ) \sin (\phi _0) \sin (\Omega _0 \phi) \sin (\Omega _0 \phi _0) \Omega _0^2-8 {w_1} \\
 \cos (\Omega _0 \phi) \cos (\Omega _0 \phi _0) \sin (\phi ) P_{{x0}} \Omega _0^2-8 {w_2} \cos (\phi ) \cos (\Omega _0 \phi _0) \sin (\Omega _0 \phi) P_{{x_0}} \Omega _0^2+8 {w_2}\\
  \cos (\phi )\cos (\Omega _0 \phi) \sin (\Omega _0 \phi _0) P_{{x_0}} \Omega _0^2 -8 {w_1} \sin (\phi ) \sin (\Omega _0 \phi) \sin (\Omega _0 \phi _0) P_{{x_0}} \Omega _0^2+8 {w_2}\\
   \cos (\phi ) \cos (\Omega _0 \phi) \cos (\Omega _0 \phi _0) P_{{y_0}} \Omega _0^2-8 {w_1} \cos (\Omega _0 \phi _0) \sin (\phi ) \sin (\Omega _0 \phi) P_{{y_0}} \Omega _0^2+8 {w_1}\\ 
   \cos (\Omega _0 \phi) \sin (\phi ) \sin (\Omega _0 \phi _0) P_{{y_0}} \Omega _0^2+8 {w_2} \cos (\phi ) \sin (\Omega _0 \phi) \sin (\Omega _0 \phi _0) P_{{y_0}} \Omega _0^2+4 m {w_1} {w_2}\\
\end{split}
\end{equation*}
\begin{equation}
\begin{split}
 \Delta  \omega  \cos (\phi ) \cos (\Omega _0 \phi) \cos (\Omega _0 \phi _0) \sin (\phi _0) \Omega _0-4 m w_1^2 \Delta  \omega  \cos (\Omega _0 \phi _0) \sin (\phi ) \sin (\phi _0) \\
 \sin (\Omega _0 \phi) \Omega _0+4 m w_1^2 \Delta  \omega  \cos (\Omega _0 \phi) \sin (\phi ) \sin (\phi _0) \sin (\Omega _0 \phi _0) \Omega _0+4 m {w_1} {w_2} \Delta  \omega \\  
 \cos (\phi ) \sin (\phi _0) \sin (\Omega _0 \phi) \sin (\Omega _0 \phi _0) \Omega _0+4 {w_1} \cos (\Omega _0 \phi) \cos (\Omega _0 \phi _0) \sin (\phi ) P_{{x_0}}+4 {w_2} \\
 \cos (\phi )\cos (\Omega _0 \phi _0) \sin (\Omega _0 \phi) P_{{x_0}}-4 {w_2} \cos (\phi ) \cos (\Omega _0 \phi) \sin (\Omega _0 \phi _0) P_{{x_0}}+4 {w_1} \sin (\phi ) \\
  \sin (\Omega _0 \phi) \sin (\Omega _0 \phi _0) P_{{x_0}}-4 {w_2} \cos (\phi ) \cos (\Omega _0 \phi) \cos (\Omega _0 \phi _0) P_{{y_0}}+4 {w_1} \cos (\Omega _0 \phi _0)\\
   \sin (\phi ) \sin (\Omega _0 \phi) P_{{y_0}}-4 {w_1} \cos (\Omega _0 \phi) \sin (\phi ) \sin (\Omega _0 \phi _0) P_{{y_0}}-4 {w_2} \cos (\phi ) \sin (\Omega _0 \phi) \\
   \sin (\Omega _0 \phi _0) P_{{y_0}}+2 m (w_1^2-w_{2}^2) \Delta  \omega  \sin (2 \phi ) (\Omega _0^2-1)+4 \cos (\phi _0) (-m w_1^2 \Delta  \omega  \sin (\phi )\\
    \sin (\Omega _0 \phi) \sin (\Omega _0 \phi _0) \Omega _0^4+m w_1^2 \Delta  \omega  \cos (\phi ) \cos (\Omega _0 \phi) \sin (\Omega _0 \phi _0) \Omega _0^3+2 m w_1 w_2 \Delta  \omega  \\
    \sin (\phi ) \sin (\Omega _0 \phi) \sin (\Omega _0 \phi _0) \Omega _0^3-m \Delta  \omega  \cos (\Omega _0 \phi) \cos (\Omega _0 \phi _0) \sin (\phi )      
  (w_2-w_1 \Omega _0){}^2 \Omega _0^2\\
  -2 m {w_1}{w_2} \Delta  \omega  \cos (\phi ) \cos (\Omega _0 \phi) \sin (\Omega _0 \phi _0) \Omega _0^2-m w_2^2\Delta  \omega  \sin (\phi ) \sin (\Omega _0 \phi) \sin (\Omega _0 \phi _0) \\
  \Omega _0^2-m \Delta  \omega  \cos (\phi ) \sin ((\phi -\phi _0) \Omega _0) ({w_2} (\Omega _0^2+1){w_1}-w_2^2 \Omega _0-w_1^2\Omega _0) \Omega _0^2-m \Delta  \omega   \\      
\cos (\phi ) \cos (\Omega _0 \phi _0) \sin (\Omega _0 \phi) ({w_2}-{w_{1}} \Omega _0){}^2 \Omega _0+m {w_2}^2 \Delta  \omega  \cos (\phi ) 
\cos (\Omega _0 \phi) \sin (\Omega _0 \phi _0)\\
\Omega _0-m \Delta  \omega  \cos ((\phi -\phi _0) \Omega _0) \sin (\phi )      
({w_2} (\Omega _0^2+1){w_1}-{w_2}^2 \Omega _0-{w_1}^2 \Omega_0)
 \Omega _0)
\end{split}
\end{equation}
\begin{equation*}
\begin{split}
\gamma(\phi)= \gamma_{0} + \frac{1}{4 \omega (\Omega _0^2-1)^2}(k^2 \Delta  (4 m {w_2}^2 \Delta  \omega  \cos (\phi ) \cos (\Omega _0 \phi) \cos (\Omega _0 \phi _0) \sin (\phi _0) \Omega _0^4-4 m {w_1} \\
{w_2} \Delta  \omega  \cos (\Omega _0 \phi _0) \sin (\phi ) \sin (\phi _0) \sin (\Omega _0 \phi) \Omega _0^4+4 m {w_1} {w_2} \Delta  \omega  \cos (\Omega _0 \phi) \sin (\phi ) \sin (\phi _0) \\
\sin (\Omega _0 \phi _0) \Omega _0^4+4 m {w_2}^2 \Delta  \omega  \cos (\phi ) \sin (\phi _0) \sin (\Omega _0 \phi) \sin (\Omega _0 \phi _0) \Omega _0^4+4 {w_1} \cos (\Omega _0 \phi) \\
\cos (\Omega _0 \phi _0)\sin (\phi ) P_{{x0}} \Omega _0^4+4 {w_2} \cos (\phi ) \cos (\Omega _0 \phi _0) \sin (\Omega _0 \phi) P_{{x0}} \Omega _0^4-4 {w_2} \cos (\phi ) \cos (\Omega _0 \phi)\\ 
\sin (\Omega _0 \phi _0) P_{{x0}} \Omega _0^4+4 {w_1} \sin (\phi ) \sin (\Omega _0 \phi) \sin (\Omega _0 \phi _0) P_{{x0}} \Omega _0^4-4 {w_2} \cos (\phi ) \cos (\Omega _0 \phi) \\
\cos (\Omega _0 \phi _0) P_{{y0}} \Omega _0^4+4 {w_1} \cos (\Omega _0 \phi _0) \sin (\phi ) \sin (\Omega _0 \phi) P_{{y0}} \Omega _0^4-4 {w_1}\cos (\Omega _0 \phi) \sin (\phi ) \\
\sin (\Omega _0 \phi _0) P_{{y0}} \Omega _0^4-4 {w_2} \cos (\phi ) \sin (\Omega _0 \phi) \sin (\Omega _0 \phi _0) P_{{y0}} \Omega _0^4 -4 m {w_1} {w_2} \Delta  \omega  \cos (\phi ) \\
\cos (\Omega _0 \phi) \cos (\Omega _0 \phi _0) \sin (\phi _0) \Omega _0^3+4 m {w_1}^2 \Delta  \omega  \cos (\Omega _0 \phi _0) \sin (\phi ) \sin (\phi _0) \sin (\Omega _0 \phi) \\
\Omega _0^3-4 m {w_1}^2 \Delta  \omega  \cos (\Omega _0 \phi) \sin (\phi ) \sin (\phi _0) \sin (\Omega _0 \phi _0) \Omega _0^3 -4 m {w_1} {w_2} \Delta  \omega  \cos (\phi )\\
 \sin (\phi _0) \sin (\Omega _0 \phi) \sin (\Omega _0 \phi _0) \Omega _0^3-4 m {w_2}^2 \Delta  \omega  \cos (\phi ) \cos (\Omega _0 \phi) \cos (\Omega _0 \phi _0) \sin (\phi _0) \Omega _0^2)\\
 +4 m {w_1} {w_2} \Delta  \omega  \cos (\Omega _0 \phi _0) \sin (\phi ) \sin (\phi _0) \sin (\Omega _0 \phi) \Omega _0^2-4 m {w_1} {w_2} \Delta  \omega  \cos (\Omega _0 \phi) \sin (\phi ) \\
 \sin (\phi _0) \sin (\Omega _0 \phi _0) \Omega _0^2-4 m {w_2}^2 \Delta  \omega  \cos (\phi ) \sin (\phi _0) \sin (\Omega _0 \phi) \sin (\Omega _0 \phi _0) \Omega _0^2-8 {w_1} \\
 \cos (\Omega _0 \phi) \cos (\Omega _0 \phi _0) \sin (\phi ) P_{{x0}} \Omega _0^2-8 {w_2} \cos (\phi ) \cos (\Omega _0 \phi _0) \sin (\Omega _0 \phi) P_{{x0}} \Omega _0^2+8 {w_2}\\
\end{split}
\end{equation*}
\begin{equation*}
\begin{split}
 \cos (\phi )\cos (\Omega _0 \phi) \sin (\Omega _0 \phi _0) P_{{x0}} \Omega _0^2 -8 {w_1} \sin (\phi ) \sin (\Omega _0 \phi) \sin (\Omega _0 \phi _0) P_{{x0}} \Omega _0^2+8 {w_2}\\
 \cos (\phi ) \cos (\Omega _0 \phi) \cos (\Omega _0 \phi _0) P_{{y0}} \Omega _0^2-8 {w_1} \cos (\Omega _0 \phi _0) \sin (\phi ) \sin (\Omega _0 \phi) P_{{y0}} \Omega _0^2+8 {w_1}\\ 
 \cos (\Omega _0 \phi) \sin (\phi ) \sin (\Omega _0 \phi _0) P_{{y0}} \Omega _0^2+8 {w_2} \cos (\phi ) \sin (\Omega _0 \phi) \sin (\Omega _0 \phi _0) P_{{y0}} \Omega _0^2+4 m {w_1} {w_2}\\
 \Delta  \omega  \cos (\phi ) \cos (\Omega _0 \phi) \cos (\Omega _0 \phi _0) \sin (\phi _0) \Omega _0-4 m {w_1}^2 \Delta  \omega  \cos (\Omega _0 \phi _0) \sin (\phi ) \sin (\phi _0) \\
 \sin (\Omega _0 \phi) \Omega _0+4 m {w_1}^2 \Delta  \omega  \cos (\Omega _0 \phi) \sin (\phi ) \sin (\phi _0) \sin (\Omega _0 \phi _0) \Omega _0+4 m {w_1} {w_2} \Delta  \omega \\  
 \cos (\phi ) \sin (\phi _0) \sin (\Omega _0 \phi) \sin (\Omega _0 \phi _0) \Omega _0+4 {w_1} \cos (\Omega _0 \phi) \cos (\Omega _0 \phi _0) \sin (\phi ) P_{{x0}}+4 {w_2} \\
 \cos (\phi )\cos (\Omega _0 \phi _0) \sin (\Omega _0 \phi) P_{{x_0}}-4 {w_2} \cos (\phi ) \cos (\Omega _0 \phi) \sin (\Omega _0 \phi _0) P_{{x0}}+4 {w_1} \sin (\phi ) \\
 \sin (\Omega _0 \phi) \sin (\Omega _0 \phi _0) P_{{x0}}-4 {w_2} \cos (\phi ) \cos (\Omega _0 \phi) \cos (\Omega _0 \phi _0) P_{{y0}}+4 {w_1} \cos (\Omega _0 \phi _0)\\
 \sin (\phi ) \sin (\Omega _0 \phi) P_{{y0}}-4 {w_1} \cos (\Omega _0 \phi) \sin (\phi ) \sin (\Omega _0 \phi _0) P_{{y0}}-4 {w_2} \cos (\phi ) \sin (\Omega _0 \phi) \\
  \sin (\Omega _0 \phi _0) P_{{y0}}+2 m ({w_1}^2-{w_2}^2) \Delta  \omega  \sin (2 \phi ) (\Omega _0^2-1)+4 \cos (\phi _0) (-m {w_1}^2 \Delta  \omega  \sin (\phi )\\
 \sin (\Omega _0 \phi) \sin (\Omega _0 \phi _0) \Omega _0^4+m {w_1}^2 \Delta  \omega  \cos (\phi ) \cos (\Omega _0 \phi) \sin (\Omega _0 \phi _0) \Omega _0^3+2 m {w_1} {w_2} \Delta  \omega  \\
 \sin (\phi ) \sin (\Omega _0 \phi) \sin (\Omega _0 \phi _0) \Omega _0^3-m \Delta  \omega  \cos (\Omega _0 \phi) \cos (\Omega _0 \phi _0) \sin (\phi )      
 ({w_2}-{w_1} \Omega _0){}^2 \Omega _0^2\\
\end{split}
\end{equation*}
\begin{equation*}
\begin{split}
  -2 m {w_1}{w_2} \Delta  \omega  \cos (\phi ) \cos (\Omega _0 \phi) \sin (\Omega _0 \phi _0) \Omega _0^2-m {w_2}^2\Delta  \omega  \sin (\phi ) \sin (\Omega _0 \phi) \sin (\Omega _0 \phi _0) \\
  \Omega _0^2-m \Delta  \omega  \cos (\phi ) \sin ((\phi -\phi _0) \Omega _0) ({w_2} (\Omega _0^2+1){w_1}-{w_2}^2 \Omega _0-{w_1}^2\Omega _0) \Omega _0^2-m \Delta  \omega  \\
\end{split}
\vspace{-18 cm}
\end{equation*}
\begin{equation}
\begin{split}     
 \cos (\phi ) \cos (\Omega _0 \phi _0) \sin (\Omega _0 \phi) ({w_2}-{w_1} \Omega _0){}^2 \Omega _0+m {w_2}^2 \Delta  \omega  \cos (\phi ) 
\cos (\Omega _0 \phi) \sin (\Omega _0 \phi _0)\\
\Omega _0-m \Delta  \omega  \cos ((\phi -\phi _0) \Omega _0) \sin (\phi )      
({w_2} (\Omega _0^2+1){w_1}-{w_2}^2 \Omega _0-{w_1}^2 \Omega_0)
 \Omega _0)
\end{split}
\end{equation}
Taking L.H. rule into the above expressions we get
$$x(\phi)= \frac{1}{4 m \Delta \omega}[4 P_{y0} \Omega_0 - 2 m \Delta \omega (w_1 -w_2 \Omega)(\phi-\phi_0)\cos\phi - m \Delta \omega (w_1 + w_2 \Omega_0) \sin \phi$$
\begin{equation}
 - 4 P_{y0} \Omega_0 \cos(\phi-\phi_0)-m \Delta \omega (w_1 + w_2 \Omega_0)\sin (\phi-2\phi_0)+4 P_{x0}\sin(\phi-\phi_0)]
\end{equation}
$$y(\phi)=\frac{1}{4 m \Delta \omega}[-4 P_{x0} \Omega-m \Delta (\phi-\phi_0)\omega(w_2-w_1\Omega_0)\sin\phi + m \Delta \Omega (w_2 + w_1 \omega_0)\cos\phi$$

\begin{equation}
 -(m \Delta \omega (w_2 + w_1 \Omega)-4 P_{x0}\Omega_0)\cos(\phi-\phi_0) + 2 P_{y0}\sin(\phi-\phi_0)]
\end{equation}
\begin{equation}
\begin{split}
z(\phi) = \frac{1}{96 m \Delta \omega} (2(\phi-\phi_{0})(48 P_{z0} +   k m \Delta \omega ({w_1}^{2}(-3+2(\phi-\phi_{0})^2)+{w_2}^2(-3 + 2(\phi-\phi_{0})^{2} ) \\ - 2 w_{1} w_{2}
 \Omega_{0}(3 +2(\phi-\phi_{0})^{2})))+6 k m ({w_{1}}^{2}-{w_{2}}^{2}) \Delta (\phi-\phi_{0}) \omega \cos{2\phi}  + 12 k P_{y0}(w_{2} + w_{1} \Omega)\\ 
 \cos(2 \phi-\phi_{0})  - 2 k (-2 k (-2 P_{x0}(\phi-\phi_{0})(w_{1} + w_{2} \Omega)+ P_{y0}(w_{2}(1 + 2 (\phi-\phi_{0})^{2})+ w_{1} \Omega_{0}(1- \\
 2(\phi-\phi_{0})^{2})))\cos\phi_{0}  -12 k m ({w_{1}}^{2}-{w_{2}}^{2})
\Delta (\phi-\phi_{0})\omega \cos{2 \phi_{0}} + 3 k m ({w_{1}}^{2}-{w_{2}}^{2}) \Delta \omega \sin {2\phi}+ \\
3 k m \Delta\omega (w_{1} + w_{2} \Omega)^{2} \sin{2(\phi-\phi0)} -12 k P_{x0}(w_{1} + w_{2} \Omega_{0})\sin(2 \phi-\phi_{0}) 
+ 12 k (2 P_{y0}(\phi-\phi_{0})\\
(w_{2} + w_{1} \Omega) + P_{x0}(w_{1}(1+2 (\phi - \phi_{0})^2)  +\Omega w_{2}(1-2(\phi-\phi_{0})^{2})))\sin\phi_{0}- 3 k m ({w_{1}}^{2}-{w_{2}}^{2}\\
\Delta(1 + 2(\phi-\phi_{0})^{2})\omega \sin 2\phi_{0})) 
\end{split}
\end{equation}

$$ p_x(\phi)=-\frac{1}{4}[m \omega \Delta (3 w_1 -w_2 \Omega_0)\cos\phi + m \Delta \omega (w_1 + w_2 \Omega_0)\cos(\phi-2\phi_0)-2(2 P_{x0} \cos $$
 \begin{equation}
(\phi-\phi_0)+ m  \Delta (\phi-\phi_0) \omega (w_1 -w_2 \Omega_0)\sin \phi0 + 2 P_{y0} \Omega_0 \sin (\phi-\phi_0))]
\end{equation}
$$p_y(\phi)= [- 2 m \Delta (\phi-\phi_0)\omega (w_2-w_1 \Omega_0)\cos \phi + 4 P_{y0} \cos(\phi-\phi_0)- 3 m w_2 \Delta \Omega \sin\phi + $$
\begin{equation}
  m w_1 \Delta \omega \Omega_0 \sin \phi + m \Delta \omega (w_2 + w_1 \Omega_0)\sin(\phi-2\phi_0)- 4 P_{x0}\Omega_0 \sin (\phi-\phi_0)]
\end{equation}
\begin{equation}
\begin{split}
 p_z (\phi)=\frac{1}{48}[48 P_{z0}+ 4 k m \Delta (\phi-\phi_0)^2 \omega (w_1 -w_2 \Omega_0)^2 + k m \Delta \omega (w_1^2 (-3 + 2(\phi-\phi_0)^2 )\\+ w_2^2(-3 + 2(\phi-\phi_0)^2 )
- 2 w_1 w_2 \Omega_0(-3 + 2(\phi-\phi_0)^2 ))+ 6 k m (w_1^2 -w_2 ^2)\Delta \omega \\ \cos 2\phi + 3 k m \Delta \omega (w_1 + w_2 \Omega_0)^2 \cos 2(\phi-\phi_0)
-2 k P_{x0}(w_1 + w_2 \omega_0)\cos (2\phi -\phi_0) + \\12 k (-2 P_{y0}(\phi-\phi_0)(w_2 - w_1 \Omega_0) +  2 P_{x0} (w_1 + w_2 \Omega))\cos \phi - 6 k m (w_1^2 -w_2^2)\\\Delta (\phi-\phi_0)\omega \sin2\phi_0]
\end{split}
\end{equation}
\begin{equation}
\begin{split}
\gamma = \Delta + \frac{1}{48 \omega \Delta}[48 P_{z0} + 4 k m \Delta (\phi-\phi_0)^2 \omega (w_1 - w_2 \Omega)+ k m \Delta \omega (w_1^2 (-3 + 2 (\phi-\phi_0)^2)\\
+ w_2^2 (-3 + 2(\phi-\phi_0)^2)- 2 w_1 w_2 \Omega_0(-3 + 2 (\phi-\phi_0)^2)) + 6 k m (w_1^2 - w_2 ^2)\Delta \omega \cos 2\phi\\ + 3k m \Delta \omega (w_1 + w_2 \omega_0)^2 \cos 2(\phi-\phi_0)
-2 k P_{x0}(w_1 +w_2 \Omega_0)\cos (2\phi - \phi_0) + 12 k (-2\\ P_{y0} (\phi-\phi_0)(w_1-w-2 \Omega)+P_{x0}(w_1+ w_2 \omega_0 ))\cos \phi   
- 6 k m(w_1^2 - w_2 ^2) \Delta \omega \cos 2\phi_0\\ - 6 k m(w_1^2 - w_2 ^2)\ \Delta \omega(\phi-\phi_0) \sin 2\phi - 12 k P_{y0}(w_2 +w_1 \Omega_0)\sin(2\phi-\phi_0)
+ 12 k(P_{y0}\\(w_2 +w_1 \Omega_0) + 2 P_{x0}{\phi-\phi_0}(w_1 -w_2 \Omega_0))\sin \phi_0 -6 k m (w_1 ^2 -w_2 ^2) \Delta (\phi-\phi_0) \omega \sin 2 \phi_0]
\end{split}
\end{equation}
\section{APPENDIX-2}\label{aap2}
 An explicit solution for the position, momentum, and energy of a charged particle in a circularly polarized wave, in terms of the particle's position and the laboratory time, was previously found by Kong and Liu \cite{Kong}.
 The vector potential and the electric field that were used respectively are as follows:
  \begin{equation}\label{a2007}
 	\vec{A}= a_0 sin(\omega t- k z) \hat{x} - g a_0 sin(\omega t - k z) \hat{y}
 \end{equation}
  \begin{equation}\label{e2007}
 	\vec{E}=E cos(\omega t- k z) \hat{x} + E sin(\omega t -k z)\hat{y}
 \end{equation} 
For the motion of a charged particle in an EM wave and an axial magnetic field ($B_{0}$), the following relations for the particle's position and lab time as functions of the phase of the EM wave have respectively been found: 
 \begin{equation}\label{tbar}
  {\bar{z}} = {\bar{\phi}} - {\sigma} sin{\bar{\phi}}
 \end{equation}
 \begin{equation}\label{zbar}
  {\bar{t}} = {\bar{\phi}} - {\bar{\sigma}} sin{\bar{\phi}}
 \end{equation}
Where $\phi= t-  \frac{z}{c}$,
$\bar{\phi}=(\omega_{0}-g\omega)\phi+\delta$,
$\bar{t}=[(\omega_{0})+a_{6}]t-[\frac{a_{5}{\omega_{0}}}{c}+a_{6}]+(\omega_{0}-g \omega)\delta$,
$$\bar{z}=\frac{(\omega_{0}-g \omega)z}{a_{6}}-(\omega_{0}-g\omega)\frac{a_{5}}{a_{6}}+\delta, \sigma = \frac{{\omega_0 \omega_E  \sqrt{{a_1}^2 + {a_2}^2}}}{(\omega_0 -g \omega)\alpha a_6 c}, \bar{\sigma} = \frac{{\omega_0 \omega_E  \sqrt{{a_1}^2 + {a_2}^2}}}{(\omega_0 -g \omega)\alpha a_6 c} + a_6,$$
$$\alpha = \sqrt{\frac{{p_{x0}}^2+{p_{y0}}^2+{p_{z0}}^2}{m^2 c^2}+1}-\frac{p_{z0}}{m c}, \alpha_x = \frac{p_{x0}}{mc} +\frac{\Omega_0}{c}y_0,\alpha_y = \frac{p_{y0}}{mc} - g \frac{\omega_E}{\omega} +\frac{\Omega_0}{c}x_0, a_1 = -\frac{c \alpha_y}{\Omega_0}$$
$$a_2 = \frac{p_{y0}}{m \Omega_0} + \frac{c \omega_E}{\Omega_0 (\omega_0 - g \omega)}, a_3 = -\frac{p_{x0}}{m \Omega_0}, a_{4}=-\frac{c \alpha_x}{\Omega}, \delta=\frac{a_{3}}{a_{2}}, a_5 = \frac{c \alpha_x \omega_E}{(\Omega_0 - g \omega \alpha)^2},$$ 
$$a_6 = \frac{p_{z0}}{m \alpha}
 -\frac{a_2 \omega_0 \omega_E}{\alpha(\omega_0 - g \omega)}, \omega_{E}=\frac{eE}{mc}, \Omega_{0}=\frac{e B_{0}}{mc},\omega_{0}=\frac{\Omega_{0}}{\alpha}$$.

As long as ${|\sigma|} < 1$ and ${|\bar{\sigma}|} < 1$, the transcendental equations (\ref{tbar}) and (\ref{zbar}) become invertible and are respectively expressed in the following forms: 
\begin{equation}\label{phibarz}
 \bar{\phi}=\bar{z}+ \Sigma_{n=1}^{\infty}\frac{2}{n} J_n (n  \sigma) sin ({n\bar{z}})
\end{equation}
\begin{equation}\label{phibart}
 \bar{\phi}=\bar{t}+ \Sigma_{n=1}^{\infty}\frac{2}{n} J_n (n \bar{ \sigma}) sin ({n\bar{t}})
\end{equation}

Using equations (\ref{phibarz}) and (\ref{phibart}) to eliminate $\phi$ from the expressions of energy and momentum (as provided in the reference \cite{Kong}), one can express the energy and momentum of the charged particle as functions of position or laboratory time.
However, it can be shown that for the resonance condition $\omega_{0}=g\omega$,
$\lim_{{\omega_{0} \to g \omega}} \sigma = -1$ and $\lim_{{\omega_{0} \to g \omega}} \bar{\sigma} = -\infty$; which implies that the above equations are not valid for the resonance condition. Thus, the expressions given in equations (\ref{phibarz}) and (\ref{phibart}) cannot be used for calculating the energy and momentum of the charged particle as a function of position or laboratory time in the resonant case. 
In addition to this, these results are not applicable to all initial conditions in the non-resonant case. For instance, when a charged particle begins with zero initial momentum and $\omega \ne g \omega_{0}$, the condition $\sigma = -1$ is satisfied. Thus, these solutions only apply to a specific set of initial conditions in the non-resonant case.
\section{APPENDIX-3}\label{app3} 
Let us use equations (\ref{chap5_eqn20}) and (\ref{chap5_eqn21}) in equation (\ref{chap5_eqn19}) and integrate; we get,
\begin{equation}\label{pz_ro}
	p_{z} = p_{z0} - \frac{g \omega_{0}}{2}  \rho^{2}
\end{equation}	
Next, use equations (\ref{chap5_eqn20}), (\ref{chap5_eqn21}), and (\ref{pz_ro}) in the expression $\gamma =\sqrt {(\gamma \dot{\rho})^2+(\gamma \rho \dot{\theta})^{2}+{p_z}^2 +1 }$ (here dot represents the derivative with respect to time $t$). Thereafter, substitute the obtained expression for $\gamma$ in the equation $\gamma-p_{z}=\Delta$, and we get
\begin{equation}
\Delta (g \Omega_0 + 1){\rho} +2 a_{0}  \sin(\phi-g\theta) = 0
\end{equation}
 For resonance case ($g \Omega = -1$), we get
 \begin{equation}
  \sin (\phi-g \theta) =0
 \end{equation}
This results in $\dot{\phi}=-g\dot{\theta}$, which, in conjunction with the resonance condition, leads to $\dot{\theta}=-\omega_{0}/\gamma$.

\section{DATA AVAILABILITY}
The data that support the findings of this study are available from
the corresponding author upon reasonable request.
\bibliography{shivam}
%
%
%
%
%
%
%
%
%
%
%
\end{document}